\documentclass[%
 aip,
 amsmath,amssymb,
 reprint,%
]{revtex4-1}

\usepackage{graphicx}% Include figure files
\usepackage{dcolumn}% Align table columns on decimal point
\usepackage{bm}% bold math
\usepackage[english]{babel}
\usepackage[utf8]{inputenc}
\usepackage[T1]{fontenc}
\usepackage{mathptmx}
\usepackage{color}
\usepackage{siunitx}
\usepackage{color}
\usepackage{bm}
\usepackage{multirow}
\usepackage{ifthen}
\usepackage{bbm}

\graphicspath{ {images/} }

\newcommand{\lbx}{\Delta x}
\newcommand{\lbt}{\Delta t}
\newcommand{\bfE}{{\bf E}}
\newcommand{\bfr}{{\bf r}}
\newcommand{\bfu}{{\bf u}}
\newcommand{\bfv}{{\bf v}}
\newcommand{\bfci}{{\bf c}_i}
\newcommand{\Bessel}[3]{{#1_{#2}(\kappa #3)}}

\definecolor{orange}{rgb}{1,0.5,0}
\definecolor{darkgreen}{rgb}{0,0.4,0.1}

\newcommand{\highlightrevision}{false}

\ifthenelse{\equal{\highlightrevision}{true}}
{

\newcommand{\revbar}[1]{{\color{red}{\sout{#1}}}}
}
{

\newcommand{\revbar}[1]{}
}

\begin{document}

\preprint{AIP/123-QED}

\title{Lattice Boltzmann Electrokinetics simulation of nanocapacitors}

\author{Adelchi J. Asta}
\affiliation{\small Sorbonne Universit\'es, CNRS, Physico-Chimie des \'electrolytes et
Nanosyst\`emes Interfaciaux, F-75005 Paris, France}
\author{Ivan Palaia}
\affiliation{\small LPTMS, UMR 8626, CNRS, Univ. Paris-Sud, Universit\'e Paris-Saclay, 91405 Orsay, France} 
\author{Emmanuel Trizac}
\affiliation{\small LPTMS, UMR 8626, CNRS, Univ. Paris-Sud, Universit\'e Paris-Saclay, 91405 Orsay, France} 
\author{Maximilien Levesque}
\affiliation{PASTEUR, D\'{e}partement de chimie, \'{E}cole Normale
Sup\'{e}rieure, PSL University, Sorbonne Universit\'{e}, CNRS, 75005 Paris, France}
\author{Benjamin Rotenberg}
\affiliation{\small Sorbonne Universit\'es, CNRS, Physico-Chimie des \'electrolytes et
Nanosyst\`emes Interfaciaux, F-75005 Paris, France}
\affiliation{\small R\'eseau sur le Stockage Electrochimique de l'Energie
(RS2E), FR CNRS 3459, France}

\markboth{}% 
{}

\date{\today}

\begin{abstract}
We propose a method to model metallic surfaces in 
Lattice Boltzmann Electrokinetics simulations (LBE),
a lattice-based algorithm rooted in kinetic theory
which captures the coupled solvent and ion dynamics
in electrolyte solutions.
This is achieved by a simple rule to impose
electrostatic boundary conditions, 
in a consistent way with the location of
the hydrodynamic interface for stick boundary conditions.
The proposed method also provides the local charge 
induced on the electrode by the instantaneous distribution of ions under
voltage. We validate it in the low voltage regime by comparison with
analytical results in two model nanocapacitors: parallel plate and coaxial
electrodes. We examine the steady-state ionic concentrations and electric 
potential profiles (and corresponding capacitance), 
the time-dependent response of the charge on the electrodes,
as well as the steady-state electro-osmotic profiles in the presence of an
additional, tangential electric field. 
The LBE method further provides the time-dependence
of these quantities, as illustrated on the
electro-osmotic response.
While we do not consider this case in the present work, which focuses
on the validation of the method, the latter readily applies 
to large voltages between the electrodes, as well as to time-dependent
voltages. 
This work opens the way to the LBE simulation of more complex systems
involving electrodes and metallic surfaces, such as sensing devices based
on nanofluidic channels and nanotubes, or porous electrodes.
\end{abstract}

\maketitle

%%%%%%%%%%%%%%%%%%%%%%%%%%%%%%%%%%%%%%%%%%%%%%%%%%%%%%%%%%%%%%%%%%%%%%%%%%
%%%%%%%%%%%%%%%%%%%%%%%%%%%%%%%%%%%%%%%%%%%%%%%%%%%%%%%%%%%%%%%%%%%%%%%%%%
\section{Introduction}

Interfaces between metals and electrolyte solutions
play the central role in electrochemistry as well
as in many analytical chemistry techniques. 
Electrodes are also necessary to apply electric field to manipulate charged 
objects in solutions, such as colloidal particles or electrolytes.
As a result, electrode-electrolyte interfaces have been
extensively studied both experimentally and theoretically for
well over a century.
Recent technological advances have made it possible to design
experimental setups in which electrolyte solutions are confined
between electrodes separated by very small distances, down 
to a few tens or hundreds of nm, or within
carbon nanotubes which may also
exhibit partially metallic behavior\cite{blase_hybridization_1994}.
The ability to build such nanocapacitors opens the way to new
analytical strategies based on electrochemistry
with a very limited number of redox-active species,
using nanofluidic devices\cite{rassaei_hydrodynamic_2012,mathwig_pushing_2013,
lemay_single-molecule_2013} or thin layer cells\cite{sun_electrochemistry_2008},
and questions our basic understanding of coupled fluid and charge
flows, or electrokinetic phenomena, through single
nanotubes\cite{bocquet_nanofluidics_2010,siria_giant_2013,
secchi_scaling_2016,jubin_dramatic_2018}.

Significant progress has been made in the understanding of the electric
double layer (EDL) at charged or metallic interfaces since the pioneering
Gouy-Chapman-Stern theory\cite{Gouy,Chapman, Stern}. 
In recent years, simulations has become a powerful tool to
predict their structure and dynamics without the need to rely on
strong simplifying assumptions, which are generally required to obtain
analytical theoretical results.
For example, Brownian dynamics simulations allowed to investigate
the relaxation of the EDL after a charge transfer
event\cite{grun_relaxation_2004}, treating the metallic electrodes
as homogeneously charged surfaces and the solvent as a dielectric continuum.
At the atomistic level, the introduction of models allowing to perform
molecular simulation of electrodes maintained at a constant potential
(as in a perfect metal), rather than constant charge\cite{siepmann_influence_1995,
reed_electrochemical_2007}, opened the way to detailed investigations
of electrochemical interfaces. These studies showed the importance of taking the
polarization of the metal by the electrolyte into account\cite{merlet_computer_2013,
Merlet2013,limmer_charge_2013,merlet_electric_2014}. However,
the computational cost of such atomistic simulations restricts their use
to small systems (below 10~nm) and relatively concentrated electrolytes
(due to the small number of ions in such small volumes).

The dynamics of ions in the bulk and in EDLs, and in turn the charging dynamics 
of nanocapacitors, results from their thermal motion (diffusion) and their 
migration due to the local electric field they experience.
Taking these factors into account allows to provide a detailed
description of the charging dynamics in capacitors
in planar\cite{bazant_diffuse-charge_2004,janssen_transient_2018} 
or more complex (\textit{e.g.} porous) geometries\cite{biesheuvel_nonlinear_2010}.
Another process by which ions move is their 
advection by the local fluid flow, which may vanish by symmetry in some
simple cases, but cannot be neglected {\it a priori}. Together with 
the fluid flow induced by the net local charge within the EDL, this
is at the origin of the above-mentioned electrokinetic phenomena,
which have been long studied theoretically or numerically
with simulations, from molecular\cite{joly_hydrodynamics_2004,joly_liquid_2006,
yoshida_molecular_2014} to models with various levels of coarse-graining
(see \textit{e.g.} Refs.~\citenum{pagonabarraga_recent_2010} 
and~\citenum{rotenberg_electrokinetics_2013} for reviews
on multiscale simulation approaches).

Among these mesoscopic simulation approaches for electrokinetics
(such as Dissipative Particle Dynamics~\cite{smiatek_mesoscopic_2009}
or Multiparticle Collision
Dynamics\cite{ceratti_stochastic_2015,dahirel_hydrodynamic_2018}), 
Lattice-Boltzmann\cite{SucciBook} (LB) has emerged as an efficient compromise 
between the simplicity of the solvent description, based on kinetic
theory and allowing to recover proper hydrodynamic behavior,
and on the flexibility with which it can be coupled to
explicit particles or free energy models to describe complex fluids.
In the former case, Molecular Dynamics (MD) coupled to
LB was successfully used to investigate the electrokinetic effects
with charged colloids\cite{lobaskin_electrophoretic_2004,lobaskin_electrophoresis_2007},
polyelectrolytes in the bulk\cite{hickey_implicit_2010} or
grafted on surfaces\cite{hickey_electrophoretic_2013}
or their translocation through nanopores\cite{datar_electrokinetic_2017},
and more recently (and closer to the subject of the
present work) to the response of EDLs to changes in the charge of
surfaces\cite{lobaskin_diffusive-convective_2016}.

The other approach, where no explicit particles are present,
exists in different flavors, which can broadly be seen as efficient numerical
solvers of the continuous electrokinetic equations, even though
their roots on kinetic theory also provide additional information
on the dynamics of species.
In that respect, treating solvent and ions on the same footing
in a multi-component LB model\cite{marini_bettolo_marconi_charge_2012} 
is a promising approach to capture correlations due in particular to the 
discrete nature of solvent molecules and ions at this coarse-grained level,
especially under extreme confinement (comparable to molecular sizes).
For larger systems, the LB method is rather coupled to
numerical schemes to describe the evolution of ions.
Assuming their instantaneous relaxation 
(on the time scale over which the fluid evolves)
toward the Poisson-Boltzmann equilibrium,
for charged\cite{wang_lattice_2006} or
constant-potential\cite{thakore_charge_2015}
walls, does not allow investigating the relaxation of the ionic
concentration and potential profiles in the EDLs.
This requires an explicit integration of the ionic
dynamics, typically solving the Nernst-Planck equation
(described below), via finite differences/elements methods. This has for example been 
used to simulate electrokinetic effects in porous 
media\cite{hlushkou_coupled_2004,hlushkou_pore-scale_2007}
or electrochemical desalination\cite{hlushkou_numerical_2016}.

An alternative hybrid approach for the dynamics of ions coupled to
the LB method for that of the fluid makes a consistent use of the
LB lattice. Inspired by previous work based on the moment propagation
method\cite{warren_electroviscous_1997}, and extending a previous
attempt with ionic fluxes computed on the lattice
node\cite{horbach_lattice-boltzmann_2001}, Capuani \textit{et al.}
proposed a method focussing instead on the ionic fluxes
through each link connecting nodes of the lattice (via the discrete
lattice velocities)\cite{capuani_discrete_2004}. This point of view has a number
of advantages, such as strictly enforcing charge conservation in particular
at solid-liquid boundaries, and offering a statistical interpretation
which can be exploited to compute other properties such as velocity
auto-correlation functions via moment
propagation\cite{rotenberg_dispersion_2008}.
This hybrid LB/link-flux method, called Lattice Boltzmann Electrokinetics
(LBE), has been successfully used
to investigate the dynamics of charged colloids\cite{pagonabarraga_mesoscopic_2005,
capuani_lattice-boltzmann_2006,giupponi_colloid_2011,rempfer_reducing_2016,kuron_moving_2016},
charged porous media and ions in oil-water mixtures\cite{rotenberg_coarse-grained_2010} 
or binary colloidal suspensions\cite{rivas_mesoscopic_2018}.
In these systems, electrostatic boundary conditions at solid-liquid
interfaces correspond to constant charge (Neumann, \textit{i.e.} constant
normal electric field), rather than constant potential (Dirichlet).

In the present work, we show that a simple rule to
impose Dirichlet electrostatic boundary conditions allows the
simulation of systems involving metallic surfaces using
LBE simulations. Specifically, the method leads to imposing
the target potential at the location of the hydrodynamic
interface, \textit{i.e.} between the solid and liquid nodes rather than solely on the 
solid nodes. In addition, it is possible to determine the instantaneous
local charge on the electrode at virtually no additional cost.
This opens the way to the simulation of the dynamic response of
electric double layers in capacitors by following the evolution
of the ionic concentrations and potential profiles as well as the charge
of the electrodes. The LBE method naturally also captures the electrokinetic 
couplings with the solvent.
The proposed implementation of electrostatic boundary conditions
is readily applicable to arbitrary electrode geometries, just
as the bounce-back rule to impose no-slip boundary conditions.

The electrokinetic equations and the LBE algorithm are presented in 
Section~\ref{sec:method}, together with the proposed method to
impose constant-potential boundary conditions and to compute the charge
induced on the (blocking) electrode by the instantaneous distribution of ions under
voltage.
We then demonstrate the validity of the method in Section~\ref{sec:results}
by considering capacitors in two geometries, parallel plate and coaxial
electrodes, in the regime of small applied voltage, for which analytical
results are available (Debye-H\"uckel theory for the ionic concentration
and electric potential profiles, together with Stokes for the steady-state
electro-osmotic profiles). 
We also show numerical results for the transient regime for electro-osmosis
in the presence of an additional, tangential electric field, for which
no analytical results are available.
While we do not consider this case in the present work, which focuses
on the validation of the method, the latter readily applies 
to large voltages between the electrodes.

%%%%%%%%%%%%%%%%%%%%%%%%%%%%%%%%%%%%%%%%%%%%%%%%%%%%%%%%%%%%%%%%%%%%%%%%%%
%%%%%%%%%%%%%%%%%%%%%%%%%%%%%%%%%%%%%%%%%%%%%%%%%%%%%%%%%%%%%%%%%%%%%%%%%%
\section{Method}
\label{sec:method}

%%%%%%%%%%%
\subsection{Electrokinetic equations}

The canonical description of electrokinetic couplings 
in a dilute electrolyte consisting of $k$ ionic species
with valencies $z_k$ and diffusion coefficients $D_k$
in a solvent characterized by its mass density $\rho$,
dynamic viscosity $\eta$ and dielectric permittivity $\epsilon_0\epsilon_r$,
couples the Poisson-Nernst-Planck equations for the dynamics
of ions and the Navier-Stokes equation for that of the solvent.
The Nernst-Planck equation is a conservation equation for the
ionic concentrations $\rho_k$:
\begin{align}
\label{eq:NP}
\frac{\partial \rho_k}{\partial t} 
&+ \nabla \cdot \left[ \rho_k \bfu + {\bf j}_k\right] =
\nonumber \\
&\frac{\partial \rho_k}{\partial t} 
+ \nabla\cdot\left[
\rho_k \bfu - D_k\nabla\rho_k - \beta D_k z_k e \rho_k\nabla\psi
\right] = 0
\end{align}
where $\beta = 1/k_BT$ with $k_B$ the Boltzmann constant and $T$ the
temperature, $e$ is the elementary charge,
$\bfu$ is the local velocity of the fluid
and where the electrostatic potential $\psi$ satisfies 
the Poisson equation:
\begin{align}
\label{eq:Poisson}
\nabla^2\psi = -\frac{1}{\epsilon_0\epsilon_r}\rho_{el}
 = -\frac{e}{\epsilon_0\epsilon_r}\sum_k \rho_k z_k
\; .
\end{align}
The three terms in the flux defined by Eq.~\ref{eq:NP} correspond
to advection, diffusion and migration under the effect of the local
electric field $-\nabla\psi$, respectively.
The advective part depends on the local velocity $\bfu$ which
is assumed to satisfy the Navier-Stokes equation for an incompressible
fluid ($\nabla\cdot\bfu=0$):
\begin{align}
\label{eq:NS}
\rho\left(  \frac{\partial \bfu}{\partial t} + (\bfu\cdot\nabla)\bfu \right)
&= \eta \Delta\bfu - \sum_k \rho_k \nabla \mu_k + {\bf f}_V^{ext}
\end{align}
with ${\bf f}_V^{ext}$ the external force density and
the chemical potentials
$\mu_k=\mu_k^{id}+\mu_k^{ex}=k_BT\ln(\rho_k/\rho_k^0)+z_ke\psi$
include an ideal part (with $\rho_k^0$ a reference concentration)
and an excess part assumed to arise only from mean-field electrostatic
interactions. The excess part results, together with 
the applied electric field $\bfE_{app}$ when present, in
a local electric force acting on the fluid
$e(\sum_k z_k \rho_k)( - \nabla \psi+\bfE_{app})$
in Eq.~\ref{eq:NS}.

These coupled equations should be solved for prescribed
boundary conditions at solid-liquid interfaces, usually
stick (no-slip) for hydrodynamics ($\bfu=0$)
and Neumann (constant field, corresponding
to fixed surface charge density) or Dirichlet
(constant potential) for electrostatics.

At equilibrium, the ionic fluxes and fluid velocities
vanish. From Eq.~\ref{eq:NP}, the concentration profiles then
follow
Boltzmann distributions $\rho_k=\rho_k^0 e^{-z_k\beta e\psi}$.
%with $\rho_k^0$ are reference concentrations
%corresponding to regions where $\psi$ vanishes
%(which need not be part of the physical system of interest).
From Eq.~\ref{eq:Poisson}, the potential satisfies the 
Poisson-Boltzmann equation:
\begin{align}
\label{eq:PB}
\nabla^2\psi & = -\frac{e}{\epsilon_0\epsilon_r}
\sum_k \rho_k^0 z_k e^{ -z_k\beta e\psi}
\; ,
\end{align}
which can be linearized for small potentials (Debye-H\"uckel limit)
as:
\begin{align}
\label{eq:DH}
\nabla^2\psi &= \kappa^2 \psi = \frac{1}{\lambda_D^2}\psi
\; ,
\end{align}
with the Debye screening length:
\begin{align}
\label{eq:lambda}
\lambda_D &= \kappa^{-1} = \left( 4 \pi l_B \sum_k \rho_k^0 z_k^2 \right)^{-1/2}
\; ,
\end{align}
where the Bjerrum length $l_B=\frac{\beta e^2}{4\pi \epsilon_0\epsilon_r}$
is the distance at which the Coulomb interaction between 
two unit charges is equal to the thermal energy ($l_B=0.7$~nm in water
at room temperature, which corresponds to all the simulation results
shown in the rest of this work).

%%%%%%%%%%%
\subsection{Lattice Boltzmann Electrokinetics}

The Lattice-Boltzmann Electrokinetics (LBE) algorithm 
is a hybrid lattice scheme coupling the standard Lattice Boltzmann (LB)
method for the dynamics of the fluid, which captures in particular
overall mass and momentum conservation, with the link-flux method for
the evolution of its composition, in particular the diffusion, advection
and migration of the ions.
Since its introduction by Capuani \textit{et al.}~\cite{capuani_discrete_2004}
it has been used and described many times and we only recall the basics
to focus on the novelty of the present work, which is the introduction
of new electrostatic boundary conditions described in the next section.

The LB method can be derived as a discretized version of a continuous
kinetic equation for the evolution of the probability density 
function $f(\bfr,\bfv,t)$ to find a fluid particle with a velocity
$\bfv$ at position $\bfr$ at time $t$. The moments of $f$ in velocity 
space provide the hydrodynamic observables, such as the local density
$\rho(\bfr,t)=\int f(\bfr,\bfv,t){\rm d}\bfv$,
local mass flux $\rho(\bfr,t)\bfu(\bfr,t)=\int f(\bfr,\bfv,t)\bfv{\rm d}\bfv$
and local stress tensor.
The Boltzmann equation with the Bhatnagar-Gross-Krook (BGK)
collision operator is discretized consistently in space (cubic grid
with lattice spacing $\lbx$), time (with time step $\lbt$)
and velocity space with a finite set of velocities $\{\bfci\}$
with associated populations $f_i(\bfr,t)\equiv f(\bfr,\bfci,t)$
and weights $w_i$. 
Here we use the three-dimensional D3Q19 lattice~\cite{SucciBook},
with 19 velocities corresponding to 0, nearest and next-nearest
neighbors (with respective norms $0$, $\frac{\lbx}{\lbt}$
and $\sqrt{2}\frac{\lbx}{\lbt}$ and weights
$\frac{1}{3}$, $\frac{1}{18}$ and $\frac{1}{36}$)
and a lattice speed unit related to the thermal
velocity $c_s^2=\frac{k_BT}{m}=\frac{1}{3}\left(\frac{\lbx}{\lbt}\right)^2$,
with $m$ the mass of the fluid particles.

The local hydrodynamic variables are computed 
exactly from the populations as:
\begin{align}
\rho(\bfr,t)&=\sum_i w_i f_i(\bfr,t)
\;\; ; \;\;
\rho\bfu(\bfr,t)=\sum_i w_i f_i(\bfr,t)\bfci
\end{align}
and the populations evolved according to:
\begin{align}
f_i(\bfr+\bfci\lbt,t+\lbt) &= f_i(\bfr,t) 
\nonumber \\
&\quad
- \frac{\lbt}{\tau}\left[ f_i(\bfr,t)- f_i^{eq}(\bfr,t) \right]
+ F_i(\bfr,t)
\label{eq:LBpop}
\end{align}
where $\tau$ is the characteristic time for the relaxation
toward the local Maxwell-Boltzmann distribution $f_i^{eq}$
and controls the viscosity of the fluid, while $F_i(\bfr,t)$
accounts for the effect of local force density.
The latter includes external forces as well as the internal
contribution of local chemical potential gradients
(see Eq.~\ref{eq:NS}). 

The ionic concentrations are discretized on the same 
spatial grid and time steps and evolved using the
link-flux method, separating the contribution
of advection from the ones arising from the ideal
and excess chemical potential gradients, as described 
in Ref.~\citenum{capuani_discrete_2004}
to which we refer the reader for the advection part.
The contributions of chemical potential gradients
are expressed in a symmetrized form by writing the fluxes
${\bf j}_k=-D_k e^{-\beta \mu_k^{ex}}
\nabla \left[ \rho_k e^{+\beta \mu_k^{ex}} \right]$.
This leads to the update of 
amount of solutes on each node,
$n_k(\bfr,t)=\rho_k(\bfr,t)\lbx^3$,
according to:
\begin{align}
\label{eq:LF1}
%\frac{\rho_k(\bfr,t+\lbt)-\rho_k(\bfr,t)}{\lbt}\lbx^3
n_k(\bfr,t+\lbt)-n_k(\bfr,t)
&=-A_0\sum_i j_k^{i}(\bfr,t)
\end{align}
where the sum runs over discrete velocities,
$j_k^{i}$ is the contribution of each link
between $\bfr$ and $\bfr+\bfci\lbt$ to
the flux of species $k$ through the cell boundary
around node $\bfr$ and $A_0$ is a lattice-dependent geometric 
factor (equal to $1+2\sqrt{2}$ for D3Q19).
The link-fluxes are given by:
\begin{align}
\label{eq:LF2}
j_k^{i}(\bfr,t) &= -d_k 
\frac{ e ^ { -\beta \mu_k^{ex}(\bfr) } + e ^ { -\beta\mu_k^{ex}(\bfr+\bfci\lbt)} }{2}
\nonumber\\                                                    
& \quad \times \left[ \frac
%{ \rho_ k(\bfr+\bfci\lbt)  e ^ { +\beta \mu_k^{ex}(\bfr+\bfci\lbt) } 
{ n_ k(\bfr+\bfci\lbt)  e ^ { +\beta \mu_k^{ex}(\bfr+\bfci\lbt) } 
   - n_ k(\bfr)  e ^ { +\beta \mu_k^{ex}(\bfr) } } { \Delta_i } \right]
\end{align}
\noindent
with $d_k=(D_k/A_0)/(\lbx^2/\lbt)$ and $\Delta_i=||\bfci||/(\lbx/\lbt)$.
While this choice of discretization leads to spurious fluxes when the lattice spacing
is too large (large potential differences between neighboring
nodes)\cite{rempfer_reducing_2016}, this form enforces that
the ionic concentrations follow the Boltzmann distribution at equilibrium.

At each time step, the excess chemical potentials are computed from the local
electrostatic potential determined from the ionic concentrations
by solving numerically the Poisson equation
as described in the next section. 
The effect of thermodynamic forces, arising from local excess chemical
potential gradients, on the dynamics of the fluid (see Eq.~\ref{eq:NS}) is expressed from
the link-fluxes, in dimensionless units, via the term:
\begin{align}
F_i(\bfr,t) &= -\frac{c_s^2}{(\lbx/\lbt)^2} \sum_k \left[ \frac{ j_k^{i}(\bfr,t) }{d_k}
- \frac{ n_ k(\bfr+\bfci\lbt)  - n_ k(\bfr) } { \Delta_i } \right]
\end{align}
in Eq.~\ref{eq:LBpop}.

No-slip hydrodynamic boundary conditions are
enforced by the bounce-back rule, which places the interface at the mid-plane
between liquid and solid nodes~\cite{SucciBook}, while setting the link-fluxes
to zero through the corresponding links ensures the absence of leakage of ions
inside the solid.
Together with the advection of ions (see Ref.~\citenum{capuani_discrete_2004}
for more details), the link-flux and LB methods give rise to an evolution
of the ionic concentrations and fluid velocity satisfying the coupled
Poisson-Nernst-Planck and Navier-Stokes equations~\ref{eq:NP}, \ref{eq:Poisson}
and~\ref{eq:NS}.

%%%%%%%%%%%
\subsection{Imposing conducting boundary conditions}
\label{sec:SOR}

The Poisson equation~\ref{eq:Poisson} must be solved numerically
at each time step to determine the electrostatic potential $\psi(\bfr)$
from the charge distribution $\rho_{el}$ on the lattice.
Following previous implementations of the LBE algorithm, we use
the Successive Over Relaxation (SOR)
method~\cite{horbach_lattice-boltzmann_2001,capuani_discrete_2004,
rotenberg_coarse-grained_2010}, which we modify as described below
to impose constant-potential boundary conditions and to determine
the charged induced at the surface of the metal.
Introducing the reduced potential $\phi(\bfr)=\beta e\psi(\bfr)$,
the Poisson equation can be rewritten as
$\nabla^2\phi + 4\pi l_B \frac{\rho_{el}}{e}=0$. 
Then, we discretize the Laplacian using a stencil consistent
with the LB lattice, which can be derived from the Taylor expansion:
$\phi(\bfr+\bfci\lbt)\approx\phi(\bfr)+\lbt\nabla\phi\cdot\bfci
+\frac{\lbt^2}{2}\nabla\nabla\phi\bf{:}\bfci\bfci$.
Using the sum rules for the lattice, 
$\sum_i w_i = 1$, $\sum_i w_i c_{i\alpha} = 0$
and $\sum_i w_ic_{i\alpha}c_{i\beta}=c_s^2\delta_{\alpha\beta}$,
where $\delta_{\alpha\beta}$ is the Kronecker symbol (1 if $\alpha=\beta$,
0 otherwise) and $\{\alpha,\beta\}\in\{x,y,z\}$ refer to the components
of the discrete velocities, it then follows that
the Laplacian can be approximated by:
\begin{align}
\label{eq:DiscreteLaplacian}
\nabla^2\phi(\bfr) 
&= \frac{2}{c_s^2\lbt^2}
\sum_i w_i \left[ \phi(\bfr+\bfci\lbt) - \phi(\bfr) \right] .
\end{align}
In practice, starting from an initial guess of the potential
(\textit{e.g.} uniform at $t=0$ or from the potential at the previous
time step), the potential is found iteratively according to:
\begin{align}
\label{eq:SOR}
\phi_{h+1}(\bfr) &= \phi_{h}(\bfr) 
+ \omega \frac{c_s^2\lbt^2}{2}\left[ \nabla^2\phi_h(\bfr) + 4\pi l_B \frac{\rho_{el}(\bfr)}{e} \right]
\end{align}
with $\omega$ a constant (here 1.4) chosen to ensure numerical stability 
and convergence as a function of iteration $h$. 
It is straightforward to see that if
convergent, the procedure yields a solution of the Poisson equation.

Up to now, this procedure has been used successfully with charged colloids
or charged porous media, in which the charge density of the solid 
is known. Note that in general the distribution of the charge within the solid
(\textit{e.g.} localized at the interface or homogeneously)
matters if one wants to model solids with a fixed surface charge
density~\cite{obliger_numerical_2013}.
In the present work, our interest goes instead to model metallic solids with fixed
potential. The simplest solution is to update the potential
as described above in the liquid while maintaining the potential of
the solid nodes at the prescribed values $\psi_s$ 
%where the subscript $s$ emphasizes that this is restricted to the solid phase.
This is possible, but the results on the liquid side
are only accurate to first order in the lattice spacing $\Delta x$. Indeed, as mentioned, the location
of the physical interface between the solid and the liquid lies at the mid-plane
between the solid and liquid nodes, not on the last layer of solid nodes 
(the situation is more complex on curved boundaries).

\begin{figure}[!htb]
\center
\includegraphics[width=0.35\textwidth]{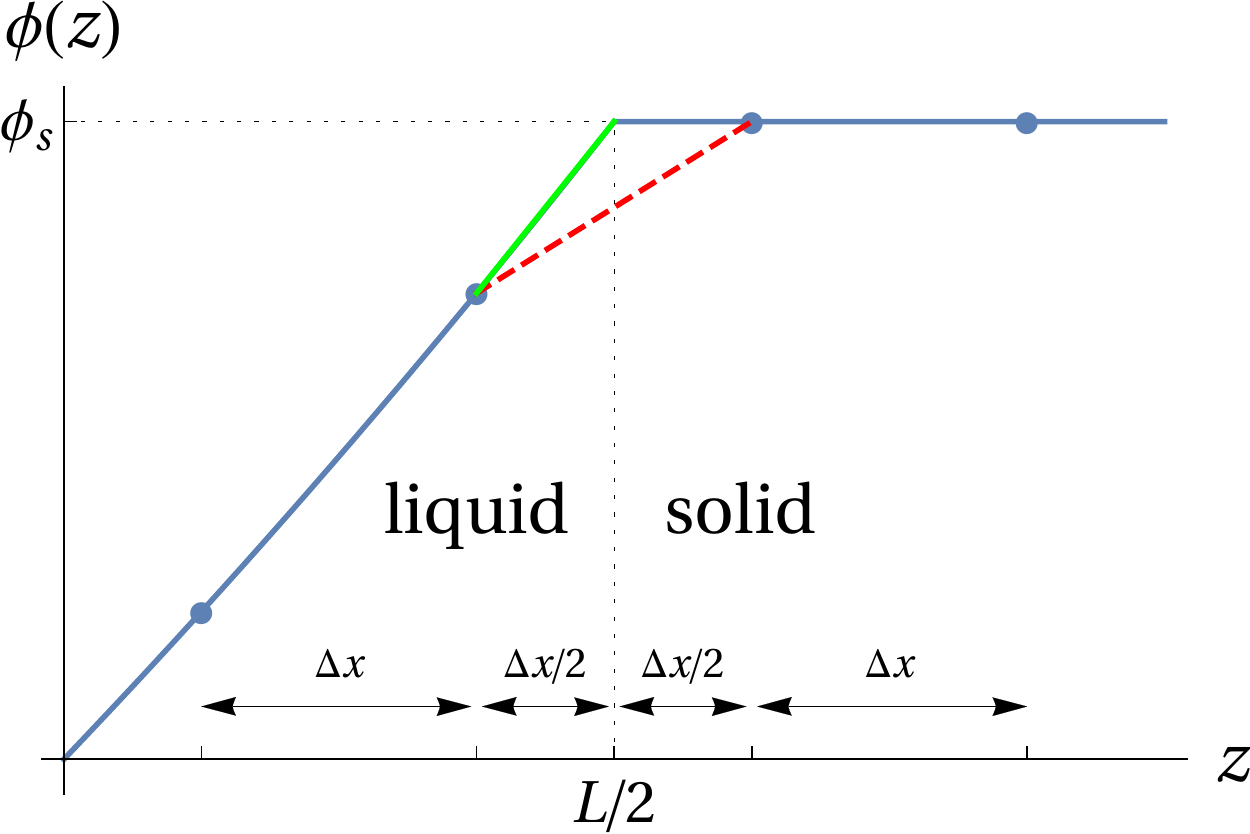}
\caption{
Enforcing the constant potential boundary condition (Dirichlet).
The electrostatic potential is displayed as a function of position, in the vicinity of a solid electrode.
For consistency with the hydrodynamic treatment, the liquid-solid interface is located halfway
between two lattice nodes, as illustrated by the vertical dotted line at $z=L/2$.
The resolution is $\Delta x$ and the reduced potential of the electrode is fixed at a constant value $\phi_s$. 
The ratio between the slopes of the thick (green) and dashed (red) lines is two. 
Unlike the former, the latter provides a poor estimation of the gradient at the interface,
as illustrated by the figure. A consistent
calculation of the gradient at the interface requires to account for this factor of two, which in turn 
leads to the modified Laplacian in Eq.~\ref{eq:DiscreteLaplacianModif} as
compared to Eq.~\ref{eq:DiscreteLaplacian}.
While $\phi$ is prescribed in the solid region, Poisson equation is solved in the liquid side.}
\label{fig:BC_ConstPot}
\end{figure}

In order to be consistent with this observation, we therefore
propose a slightly modified algorithm:
For each boundary link, \textit{i.e.} such that $\bfr$ and $\bfr+\bfci\lbt$ 
belong to different phases (interfacial nodes), we simply multiply by 2 the difference 
appearing in Eq.~\ref{eq:DiscreteLaplacian} when computing the
Laplacian in Eq.~\ref{eq:SOR} (in order to determine the potential on interfacial
liquid nodes). 
The fact that this effectively places the boundary condition 
at the mid-plane is illustrated on Figure~\ref{fig:BC_ConstPot} in the case of
a one-dimensional geometry.
A related discussion can be found in Ref.~\citenum{hlushkou_coupled_2004},
where the ion dynamics was simulated using finite elements
(see their Eq.~15 \textit{seq.}).
The proposed modification applies this idea to the stencils
used for differential operators consistent with the LB lattice
(for a discussion of stencils in the bulk, see
Ref.~\citenum{thampi_lattice-boltzmann-langevin_2011}).
It proves convenient to reformulate the modification in a compact form by introducing
the characteristic function of the solid:
\begin{align}
\chi_s(\bfr) &= 
\begin{cases}
1 & \text{if } \bfr \text{ is a solid node,}\\
0 & \text{if } \bfr \text{ is a fluid node.}
\end{cases}
\end{align}
Eq.~\ref{eq:DiscreteLaplacian} is then replaced by:
\begin{align}
\label{eq:DiscreteLaplacianModif}
\nabla^2\phi(\bfr) 
&= \frac{2}{c_s^2\lbt^2}
\sum_i w_i \left[ \phi(\bfr+\bfci\lbt) - \phi(\bfr) \right]
\times
\nonumber \\
& \; \hskip 2.5cm \left[ 1 + \chi_s(\bfr+\bfci\lbt) - \chi_s(\bfr) \right]
\end{align}
when solving the Poisson equation via Eq.~\ref{eq:SOR}.
A {\it bona fide} feature of this reformulation is that it is parametrization independent, and
can be used for arbitrary geometry of the solid electrode.
Note that this introduces a correction (with respect to
Eq.~\ref{eq:DiscreteLaplacian}) only at the boundaries,
which can be shown using the above-mentioned Taylor
expansion and sum rules to correspond to a surface term 
$2\nabla\phi(\bfr) \cdot \nabla \chi_s(\bfr)
=-\frac{\sigma}{\epsilon_0\epsilon_r}{\bf n}$,
with $\sigma$ the local surface charge density
and ${\bf n}$ the local normal unit vector pointing
out of the electrode (the factor of 2 again corresponds
to the location of the interface between the solid
and liquid nodes, as sketched in Fig. \ref{fig:BC_ConstPot}).

Once the potential distribution inside the liquid is known, in particular
at the interfacial liquid nodes, we can compute the charge $Q$ of the electrodes  
using again the Poisson equation as: 
\begin{align}
\label{eq:Qelec}
Q &= \lbx^3 \sum_{\bfr\in elec} \rho_{el}(\bfr)
= -\frac{e \lbx^3}{4\pi l_B} \sum_{\bfr\in elec} \nabla^2\phi(\bfr) 
\end{align}
where the Laplacian is computed via Eq.~\ref{eq:DiscreteLaplacianModif}
and vanishes everywhere inside the electrode except at interfacial nodes,
as expected for the charged induced by the polarization of a metal.

We will show in Section~\ref{sec:results} that the method
presented in this section allows to recover the
correct potential throughout the liquid and in turn the correct ionic 
density profiles at steady-state, as well as the corresponding capacitance
of the electrode with second order accuracy in the lattice spacing.
As for the rest of the link-flux method, the discretization of the
differential operators is only accurate for sufficiently small
variations of the considered quantities (in particular of the
potential) between neighboring nodes.
We underline however that the voltage between electrodes needs
not be small and that non-linear electrostatic regimes can be simulated
using the present method provided that the lattice spacing is 
well chosen.

%%%%%%%%%%%%%%%%%%%%%%%%%%%%%%%%%%%%%%%%%%%%%%%%%%%%%%%%%%%%%%%%%%%%%%%%%%
%%%%%%%%%%%%%%%%%%%%%%%%%%%%%%%%%%%%%%%%%%%%%%%%%%%%%%%%%%%%%%%%%%%%%%%%%%
\section{Results and discussion}
\label{sec:results}

\begin{figure}[!htb]
\center
\includegraphics[width=0.48\textwidth]{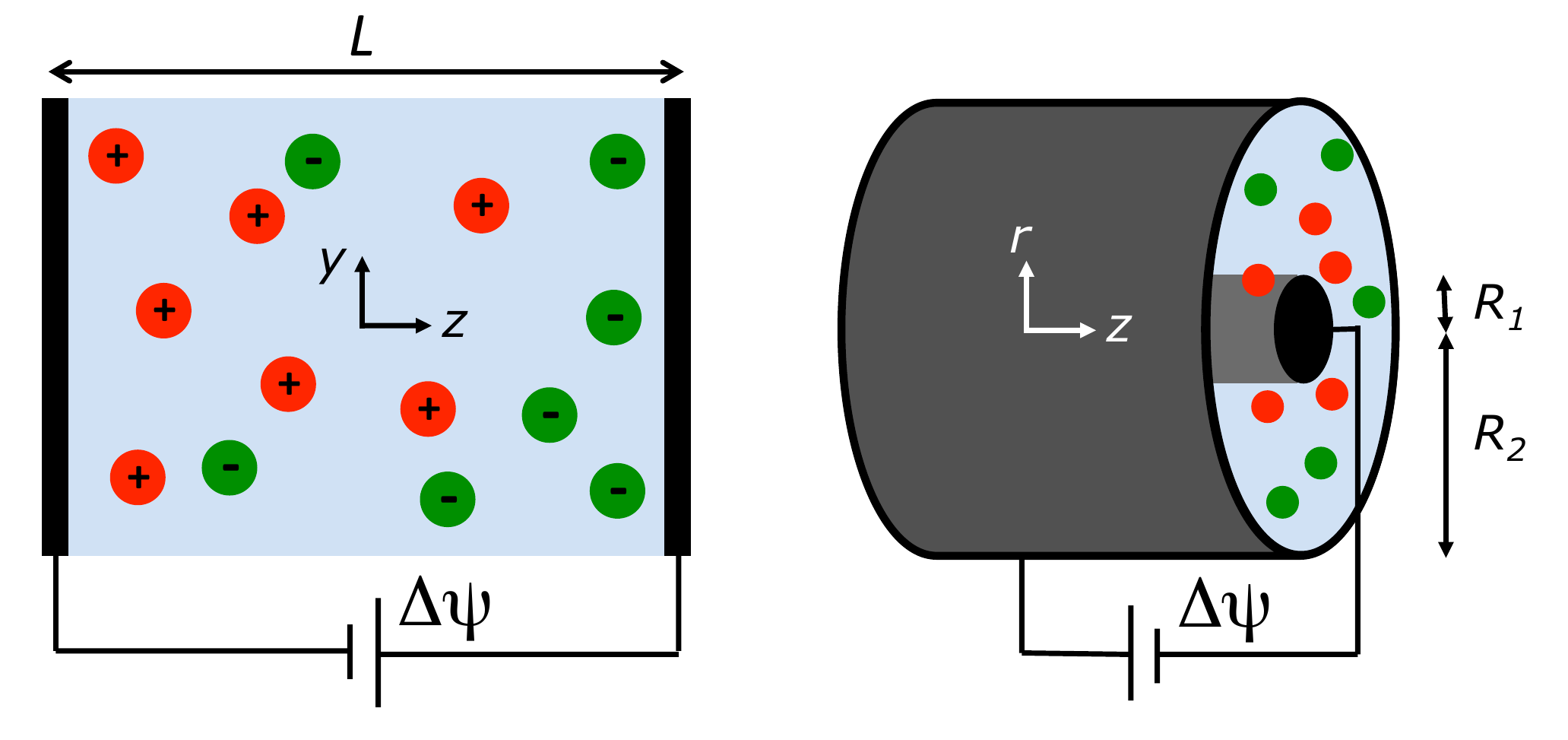}
\caption{
Capacitors consisting of an electrolyte confined between two metallic
electrodes maintained at a constant potential difference $\Delta \psi$.
Two geometries are considered: parallel plate capacitor (left), 
electrodes separated by a distance $L$ and
coaxial capacitor (right) with electrodes of inner and outer radii
$R_1$ and $R_2$.
In the following we also consider the electro-osmotic flow
induced in the charged capacitors by an additional electric field
in the $y$ (resp. $z$) direction for the parallel plate (resp. coaxial)
capacitor.
}
\label{fig:systems}
\end{figure}

In the following, we validate our approach to impose constant-potential
boundary conditions in LBE simulations by considering
cases for which it is possible to obtain analytical results, 
in the linear regime. However the method can also be readily applied 
without this restriction. 
We consider two geometries, illustrated in Fig.~\ref{fig:systems},
corresponding to parallel plate and cylindrical (coaxial) capacitors,
with a 1:1 electrolyte ($z_+=-z_-=1$) at concentration $\rho_s$
corresponding to a Debye screening length
$\lambda_D = \left( 8\pi l_B \rho_s \right)^{-1/2}$.
%\begin{align}
%\lambda_D &= 
%\left( 4\pi l_B \sum_k \rho_k z_k^2 \right)^{-1/2}
% = \left( 8\pi l_B \rho_s \right)^{-1/2}
%\end{align}
We assume for simplicity that both cations and anions have
the same diffusion coefficient $D_+=D_-=D$, 
but the simulations can be readily performed without this 
restriction.

%%%%%%%%%%%
\subsection{Parallel plate capacitor}

We first consider parallel plate capacitors with
two planar electrodes separated by a distance $L$
(in the $z$ direction, with $z=0$ at the mid-plane).
Starting from an uncharged capacitor, we apply
at $t=0$ a voltage $\Delta\psi=\psi_2-\psi_1=2.5$~mV
between the two electrodes, or in reduced
units (in terms of the thermal voltage $k_BT/e\approx25$~mV):
$\beta e \Delta\psi=0.1$. With such a small
reduced voltage, it is possible to linearize the 
Poisson-Nernst-Planck equation to obtain the 
time-dependent charge on the positive electrode $Q(t)$
as well as the steady-state potential and ionic density
profiles in the capacitor, which corresponds to the
Debye-H\"uckel (DH) theory.

LBE simulations in this geometry are performed
for a system with periodic boundary conditions in
all directions, with $N_x=N_y=1$
in the directions parallel to the surfaces
(this is sufficient to simulate infinite planar walls,
as we checked by also performing simulations for
$N_x=N_y=3$ for one of the systems). 
In the direction perpendicular to the electrodes we use 
$N_z=N_f+6$ nodes, where $N_f=L/\lbx$ (with $L$ the distance between the
solid/liquid interfaces and $\lbx$ the lattice spacing) is the number
of layers of fluid nodes, and 3 layers of nodes on each
side of the liquid for the two electrodes. This choice ensures
that there is no effect of the periodic boundary conditions
in this direction on the charged induced at the surface of
each electrode. 
We use a BGK relaxation $\tau=\lbt$, which corresponds
to a kinematic viscosity of
$\nu=\frac{\eta}{\rho}=\frac{1}{6}\frac{\lbx^2}{\lbt}$.
The diffusion coefficient of the ions is taken as 
$0.05\frac{\lbx^2}{\lbt}$, to ensure that the Schmidt
number $Sc=\nu/D$ is larger than one, as for small ions
in water (even though the order of magnitude is larger in this case).
The potentials of the two electrodes are arbitrarily 
chosen as $\psi_1=0.1~k_BT/e$ and $\psi_2=0.2~k_BT/e$ 
to apply the desired voltage, but the resulting evolution of the ionic 
densities and electrode charge do not depend on the absolute potentials, 
as expected.

%%%%%
\subsubsection{Potential and concentration profiles}

\begin{figure}[!htb]
\center
\includegraphics[width=0.45\textwidth]{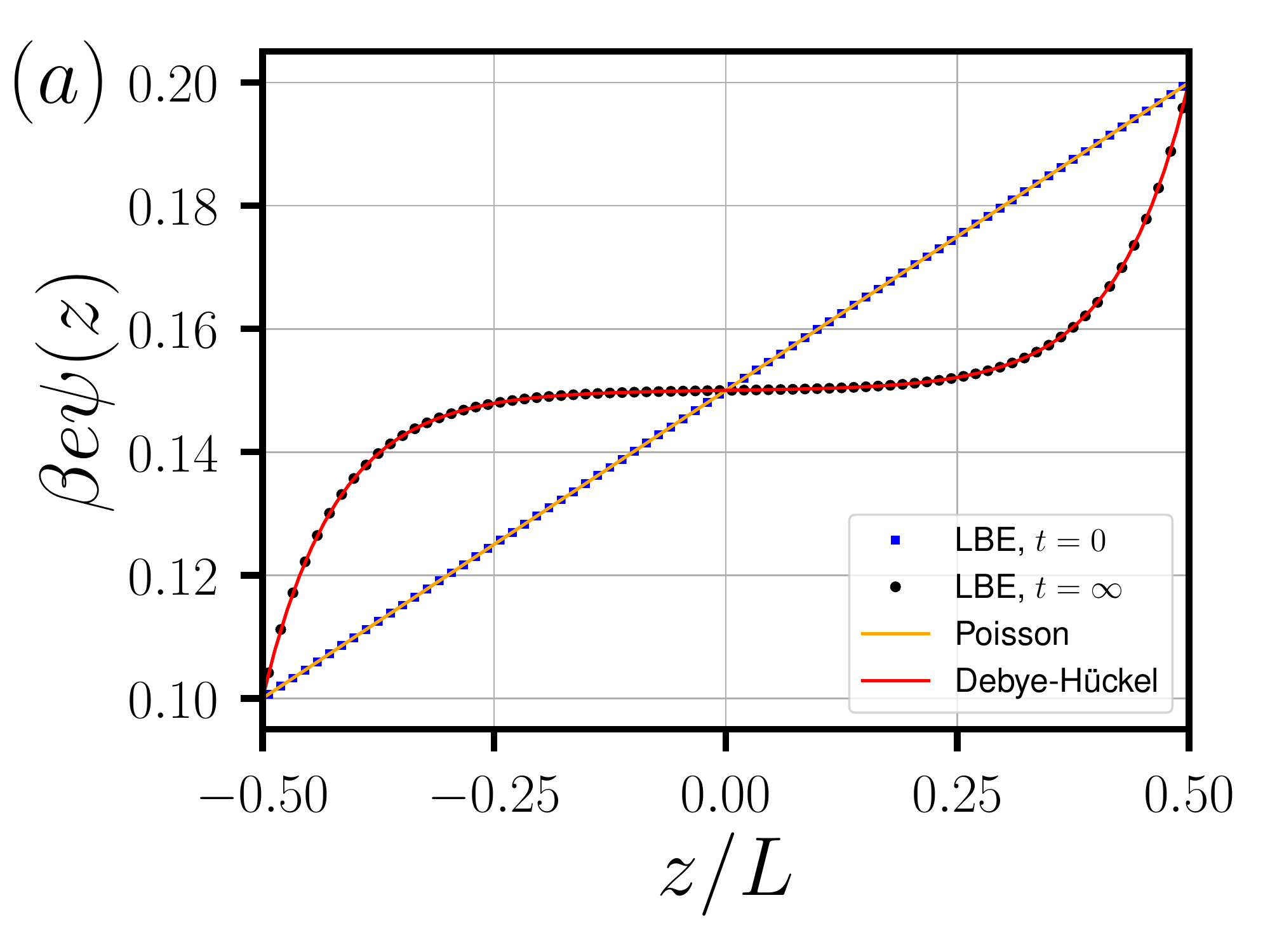}
\includegraphics[width=0.45\textwidth]{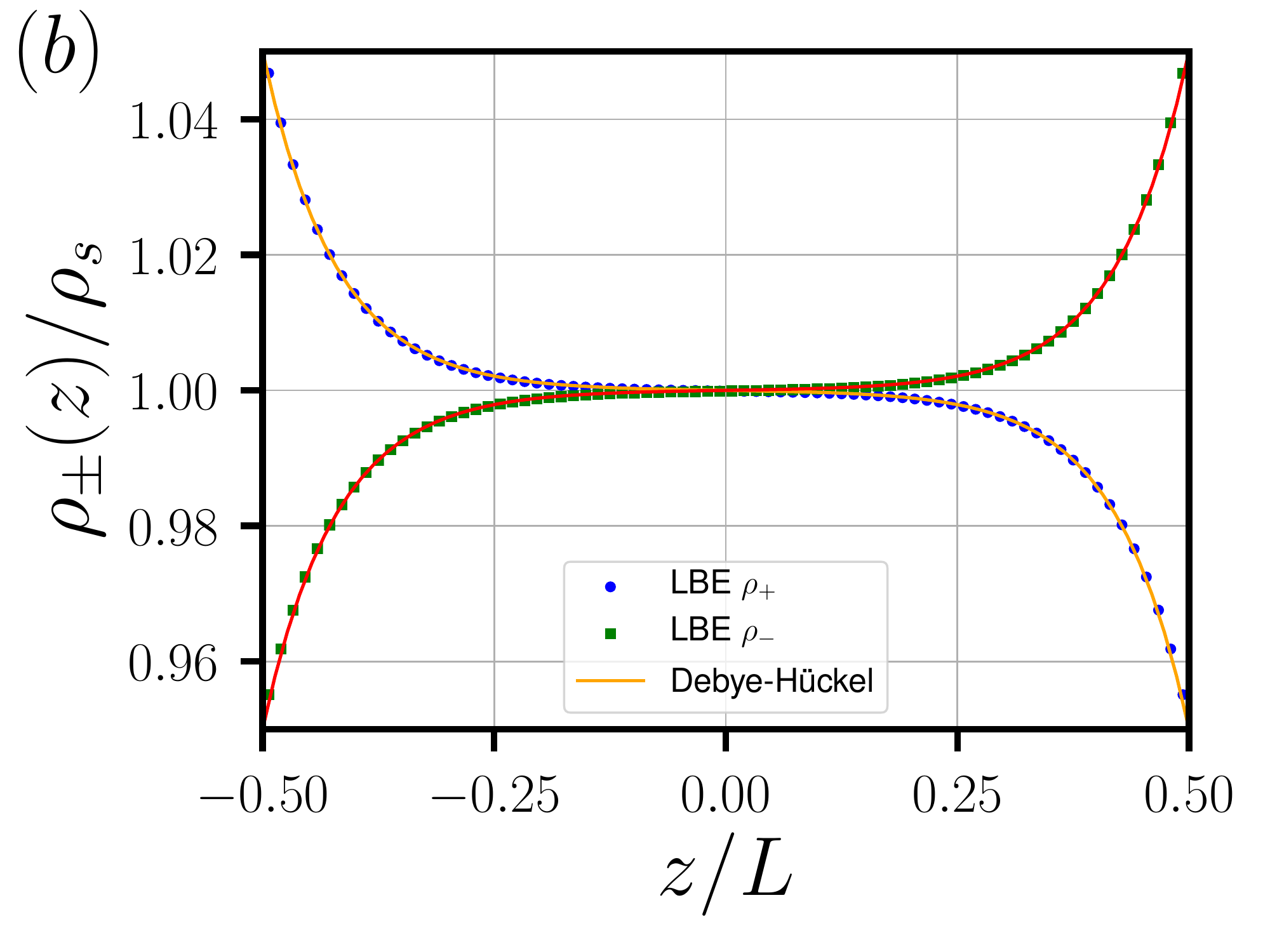}
\caption{
Steady-state electrostatic potential $\psi$ (a) and ionic concentration 
$\rho_\pm$ (b) profiles
in a parallel capacitor, obtained from Lattice-Boltzmann Electrokinetics
simulations (LBE, symbols) and Debye-H\"uckel theory (lines).
Results are normalized by the thermal potential $k_BT/e$
and salt concentration $\rho_s$, respectively.
In panel (a), we also indicate the initial potential profile:
Right after establishing the potential drop and
before the ions start to move, the fluid is neutral and the solution
of the Poisson equation in this geometry is linear, as for a
simple dielectric parallel plate capacitor.
Simulations are performed for a separation $L=76\lbx$, with a
lattice spacing $\lbx=l_B/1.44$, with $l_B$ the Bjerrum length,
a salt concentration corresponding to a Debye length $\lambda_D=6\lbx$,
and a reduced voltage $\beta e\Delta\psi=0.1$.
}
\label{fig:slit_phi_rho}
\end{figure}

Before examining the charge induced on the electrodes and the corresponding
capacitance, we first examine the potential and concentration profiles
through the capacitor, which are reported in 
Figure~\ref{fig:slit_phi_rho} for simulation
parameters indicated in its caption.
As explained above, the initial potential profile
corresponds to the solution of the Poisson equation
for a neutral capacitor, since the charge density
vanishes inside the liquid because 
$\rho_+(z)=\rho_-(z)=\rho_s$ everywhere before
the ions start moving. The corresponding initial
electric field drives the cations and anions 
toward opposite electrodes. Once the electric double
layers are established, there is no field in the bulk
part of the liquid, \textit{i.e.} at distances
much larger than $\lambda_D$ (this can be achieved
only in the regime $\lambda_D\ll L$).

The solution of the DH equation~\ref{eq:DH}
for the parallel plate capacitor with boundary conditions 
$\psi(+L/2)=\psi_2$ and $\psi(-L/2)=\psi_1$ is given by:
\begin{align}
\label{eq:DHslitpotential}
\psi^{DH}(z) &= \frac{\psi_1+\psi_2}{2}
+ \left(\frac{\psi_2 - \psi_1}{2}\right)
\times\frac{ \sinh(\kappa z) } { \sinh(\kappa L/2) }
\; .
\end{align}
Therefore in steady-state regime and the small voltage limit, both the potential and ionic density profiles decay 
exponentially from the surface, with a decay length $\lambda_D$.
The LBE results are in excellent agreement
with these analytical predictions in the considered
range of physical and simulation parameters
(which are the same as for Figure~\ref{fig:charge_capacitance}a).
This is a first validation of the proposed method to
impose the fixed potential boundary conditions.

%%%%%
\subsubsection{Charge and capacitance}

\begin{figure}[!htb]
\center
\includegraphics[width=0.45\textwidth]{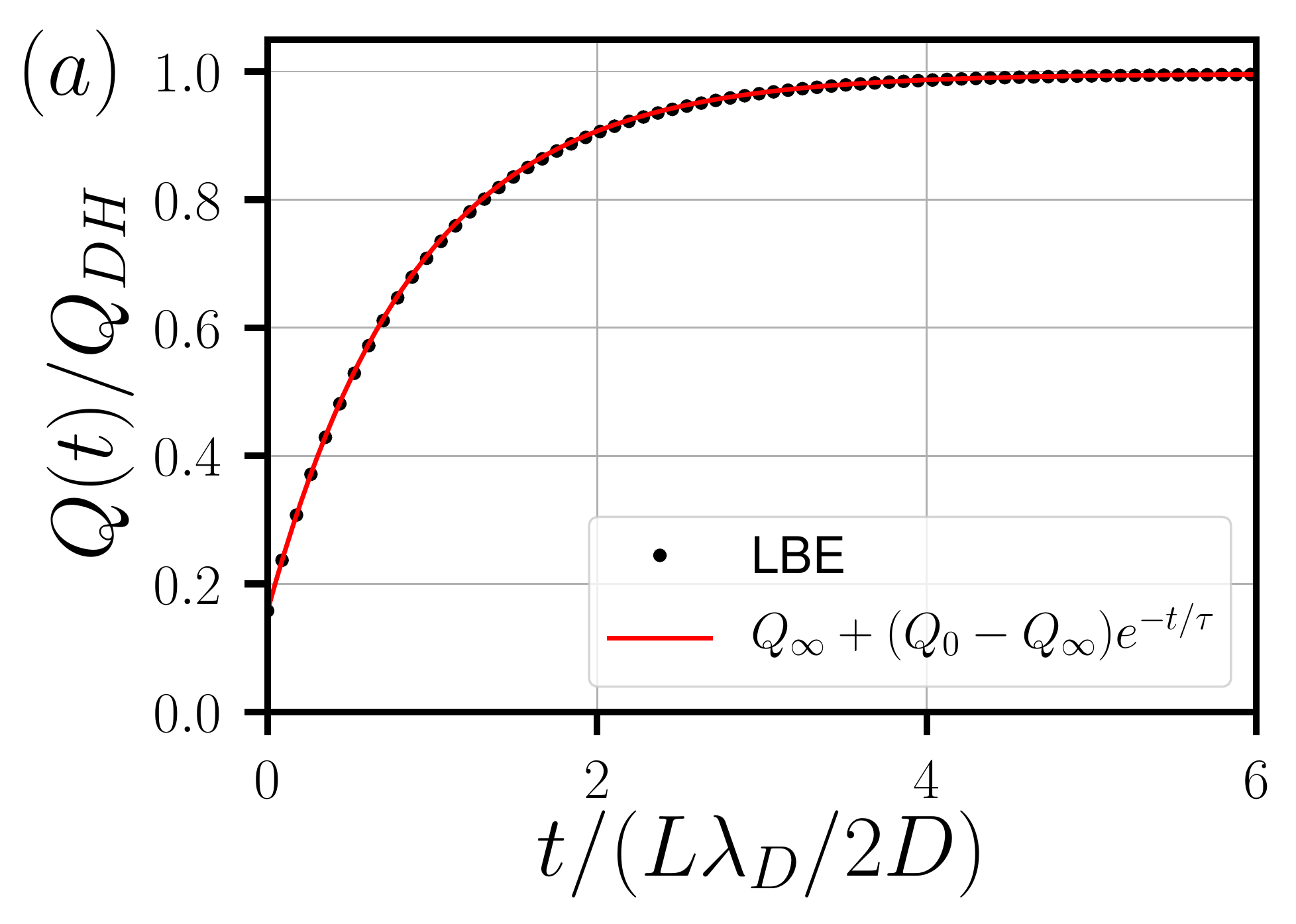}
\includegraphics[width=0.45\textwidth]{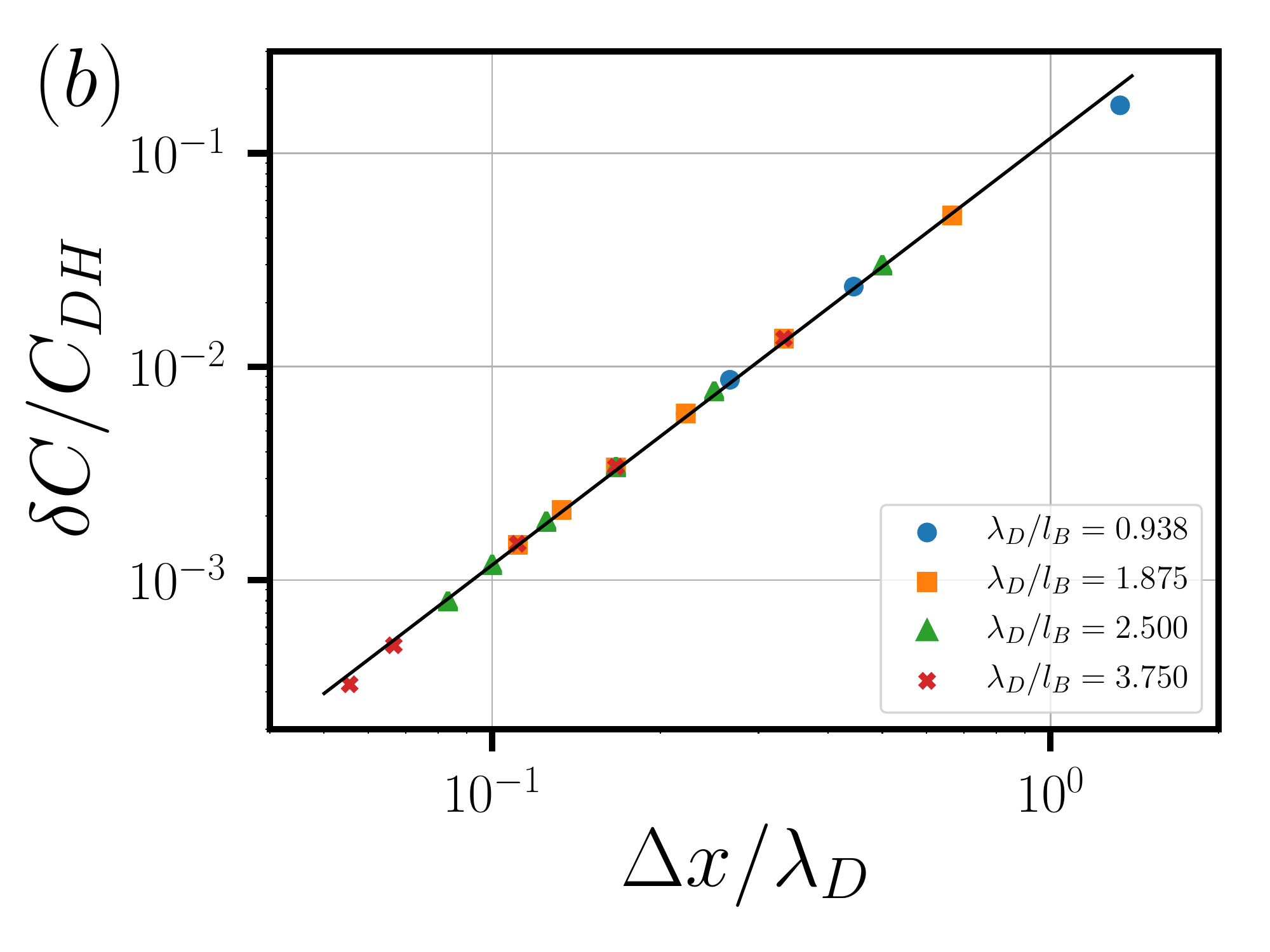}
\caption{
(a) Charging a parallel plate capacitor:
The charge obtained from Lattice Boltzmann Electrokinetics (LBE)
simulations, normalized by the Debye-H\"uckel prediction for
the surfacic capacitance $C_{DH}=\epsilon_0\epsilon_r /2\lambda_D$,
as a function of time normalized by $L\lambda_D/2D$.
The initial value of the charge coincides with the expected
value for a dielectric (neutral) capacitor $C_0=\epsilon_0\epsilon_r/L$.
Simulations are performed for a separation $L=76\lbx$, 
with a lattice spacing $\lbx=l_B/1.44$, with $l_B$ the Bjerrum length,
a salt concentration corresponding to a Debye length $\lambda_D=6\lbx$,
and a reduced voltage $\beta e\Delta\psi=0.1$.
Results are shown only every 400 steps for clarity.
The line shows an exponential fit of the LBE results
(see Figure~\ref{fig:slit_dynamics} for a discussion of the characteristic
times), while horizontal and vertical lines are only guides for the eye.
(b) Influence of the lattice spacing. 
The relative deviation of the simulated capacitance (computed from $Q_\infty$)
with respect to the Debye-H\"uckel prediction is reported as a function of
the ratio $\lbx/\lambda_D$, for several salt concentrations 
corresponding to different ratios $\lambda_D/l_B$ and a
fixed ratio $L/l_B=52.5$.
The line has a slope of 2. 
}
\label{fig:charge_capacitance}
\end{figure}

As explained in Section~\ref{sec:SOR}, we can compute the instantaneous
charge $Q(t)$ on the electrode surface from the potential distribution
(once it has been determined from the ionic concentration via the Poisson
equation) using Eq.~\ref{eq:Qelec}.
Figure~\ref{fig:charge_capacitance}a shows the charge as a function
of time for a capacitor with electrodes separated by a distance
$L\approx52.8 l_B\approx36.9$~nm 
and electrolyte concentration (0.011~mol.L$^{-1}$) such that 
$\lambda_D\approx4.2 l_B\approx 2.9$~nm.
The simulation parameters are indicated in the caption of the figure.
The charge is reported normalized by the DH prediction for the
surfacic capacitance:
\begin{align}
\label{eq:SLITcapaDH}
C_{DH}&=\epsilon_0\epsilon_r /2\lambda_D
\; ,
\end{align}
which can be interpreted physically as the capacitance for
two parallel plate capacitors with distance $\lambda_D$ in series.
Time is normalized by $L\lambda_D/2D$.
The results nicely converge to the DH prediction, which is 
expected to be valid for such a small voltage and takes the 
form of Eq. \ref{eq:SLITcapaDH} when $\lambda_D \ll L$. The charging
dynamics will be analyzed in more detail in section~\ref{sec:chargedynamics},
but one can already note the exponential form of the charge as
a function of time, illustrated by the solid line.
Another point of interest is the initial value of the charge,
which does not vanish once voltage is applied, 
but rather corresponds to the value for a dielectric (neutral) capacitor:
$C_0=\epsilon_0\epsilon_r/L$. This is due to the fact that the
liquid is neutral before the ions start moving  
(see the potential distribution inside the liquid in 
Figure~\ref{fig:slit_phi_rho}a).

Of course, the accuracy of the simulation 
results depends on the level of discretization,
more specifically the grid spacing $\lbx$ with respect to the physical
lengths. The latter are generally in the order $l_B<\lambda_D<L$,
even though the order of the last two can be reversed for small
electrolyte concentrations and distances between electrodes.
The grid spacing must be sufficiently small to resolve the electric
double layers at steady-state ($\lbx/\lambda_D<1$).

Figure~\ref{fig:charge_capacitance}b shows the relative error
on the steady-state capacitance with respect to the DH result
as a function of $\lbx/\lambda_D$,
for a fixed ratio $L/l_B=52.5$ and several values of $\lambda_D/l_B$.
The slope of 2 on this double logarithmic scale indicates
that
\begin{align}
\frac{|C_{LBE}-C_{DH}|}{C_{DH}} &\propto
\left( \frac{\lbx}{\lambda_D} \right)^2
\; ,
\end{align}
for all considered cases,
\textit{i.e.} that our algorithm to impose constant-potential boundary 
conditions and to determine the surface charge induced by the ionic 
distributions in the electrolyte is accurate to second order.
Note that we have pushed the numerical results to the
rather extreme case of $\lambda_D\approx l_B$: this is a high concentration
regime in which the DH theory itself becomes too crude an approximation,
because correlations between ions (in particular due to excluded volume)
cannot be neglected.

%%%%%
\subsubsection{Charging dynamics}
\label{sec:chargedynamics}

\begin{figure}[!htb]
\center
\includegraphics[width=0.45\textwidth]{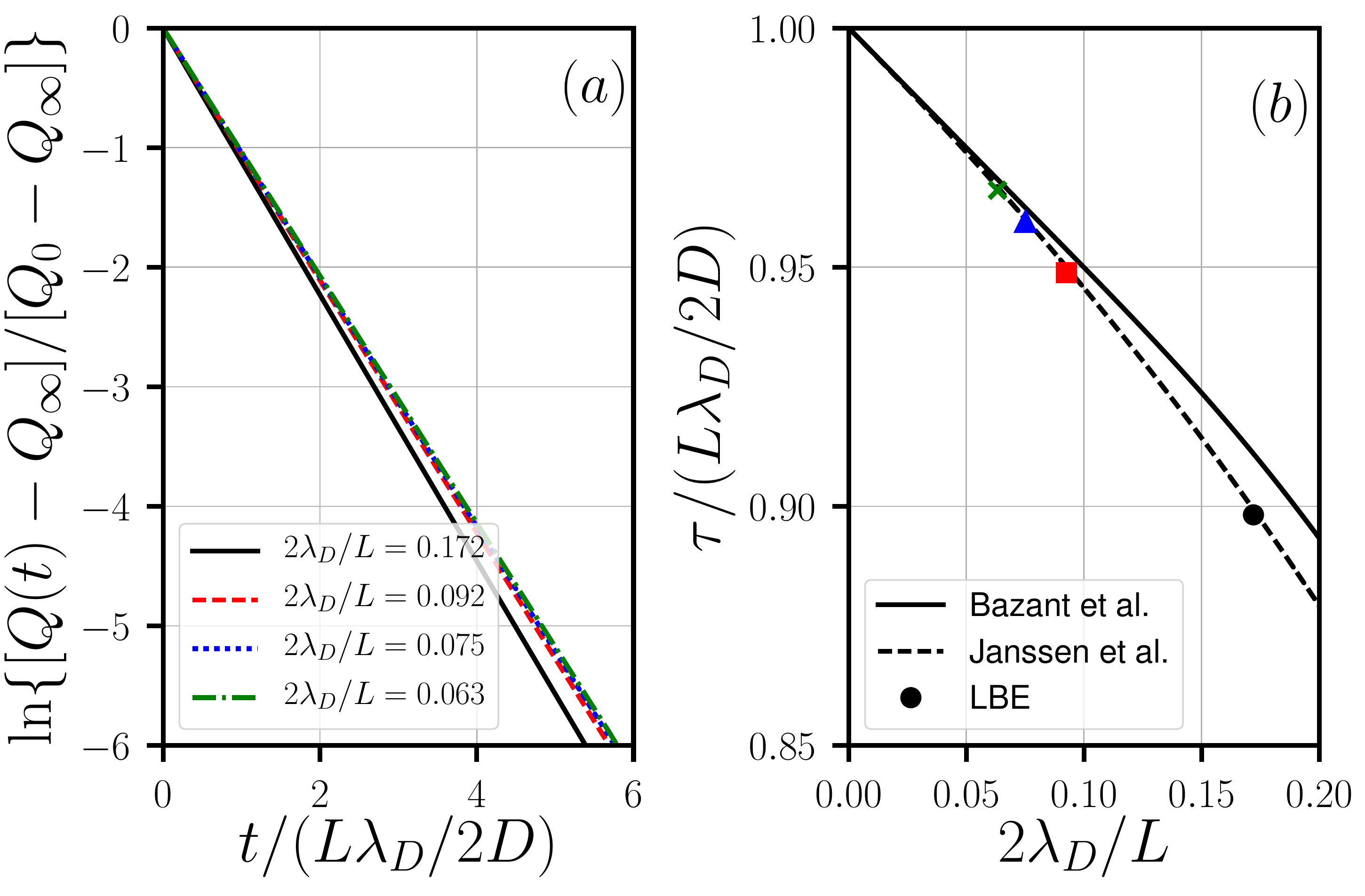}
\caption{
Charging dynamics in parallel plate capacitors.
(a) Relaxation of the charge of the electrode $Q(t)$
from its initial value $Q_0$ to its final value $Q_\infty$,
plotted on a logarithmic scale to illustrate the exponential
decay, which allows to define a relaxation time $\tau$.
(b) Relaxation time, normalized by the characteristic time
$L\lambda_D/2D$, as a function of the ratio between the Debye
screening length and the half-distance between the electrodes.
Simulations are performed for several inter-electrode
distances $L$, corresponding to the colors indicated in panel (a),  
with a lattice spacing $\lbx=l_B/4.8$, with $l_B$ the Bjerrum length,
a salt concentration corresponding to a Debye length $\lambda_D=8\lbx$,
and a reduced voltage $\beta e\Delta\psi=0.1$.
The relaxation time for each $L$ is reported in panel (b)
with the corresponding color. The simulations results
are also compared to the analytical predictions in Eq.~(36) of
Ref.~\citenum{bazant_diffuse-charge_2004} and
in Eq.~(29) and preceding definitions
of Ref.~\citenum{janssen_transient_2018}. 
}
\label{fig:slit_dynamics}
\end{figure}

The LBE simulations do not only provide the steady-state electrode charge
and potential/concentration profiles, but also their evolution with time.
Figure~\ref{fig:slit_dynamics}a reports simulation results for the 
electrode charge similar to those of Figure~\ref{fig:charge_capacitance}a, 
at fixed salt concentration (0.065~mol.L$^{-1}$,
corresponding to $\lambda_D=1.2$~nm) 
and resolution ($\lbx/l_B$) but for several distances between electrodes $L$
(see caption for details) 
and in a scale that emphasizes the exponential relaxation of $Q(t)$ 
toward the steady-state solution. This scale clearly shows that the 
corresponding characteristic time (inverse of the slope) depends 
on the system.

As pointed out \textit{e.g.} by Bazant and coworkers~\cite{bazant_diffuse-charge_2004},
the decay time is neither the Debye relaxation time $\lambda_D^2/D$,
(which is the relaxation time for bulk electrolytes)
corresponding to diffusion over the Debye length,
nor the diffusion time over the distance $L$ between the electrodes,
but rather $\sim L\lambda_D/2D$.
More accurate analytical expressions have been derived 
in Ref.~\citenum{bazant_diffuse-charge_2004} and more recently
by Janssen and Bier in Ref.~\citenum{janssen_transient_2018}, which include
a correction of order $\lambda_D^2/D$. The result can be interpreted as 
an $RC$ charging time taking into account the capacitance
of the electrode-electrolytes interfaces, estimated by $C_{DH}$,
and the resistance of the bulk electrolyte, using the 
conductivity estimated via the Nernst-Einstein expression
and considering a slab of width $\approx L-\lambda_D$ of electrolyte.
The characteristic decay time $\tau$ is reported in 
Figure~\ref{fig:slit_dynamics}b, normalized
by $L\lambda_D/2D$, as a function of the ratio $2\lambda_D/L$.
The results are in perfect agreement with the results of
Ref.~\citenum{janssen_transient_2018}, which also coincide
with that of Ref.~\citenum{bazant_diffuse-charge_2004}
for $\lambda_D\ll L$.

%%%%%%%%%%%
\subsubsection{Electrokinetic effects}
\label{sec:slit_eof}

Finally, the LBE method is able to capture the electrokinetic coupling between 
the ions and the solvent. This is illustrated in the present
case of constant-potential walls by examining the electro-osmotic response
of the \textit{charged} parallel plate capacitor (obtained as the steady-state
of the previous sections) to an additional electric field $E_y$ parallel
to the electrodes.
Note that in a real system of a capacitor with finite lateral
dimensions, such an additional field would be applied by other electrodes,
located outside of the capacitor, and the field lines would be modified
compared to the simplified case considered here for validation purposes.
For sufficiently small applied field, the electro-osmotic flow is laminar and
the steady-state solution of the Navier-Stokes equation~\ref{eq:NS}
in this geometry, with no-slip boundary conditions
and in the Debye-H\"uckel limit, is given by: 
\begin{align}
\label{eq:DHslitEOF}
u_y(z)&= \frac{\epsilon_0\epsilon_r E_y
(\psi_2 - \psi_1)}{\eta}
\times
\frac{1}{2}
\left(\frac{ \sinh(\kappa z) } { \sinh(\kappa L/2) }
- \frac{2z}{L}\right)
\end{align}

\begin{figure}[!htb]
\center
\includegraphics[width=0.45\textwidth]{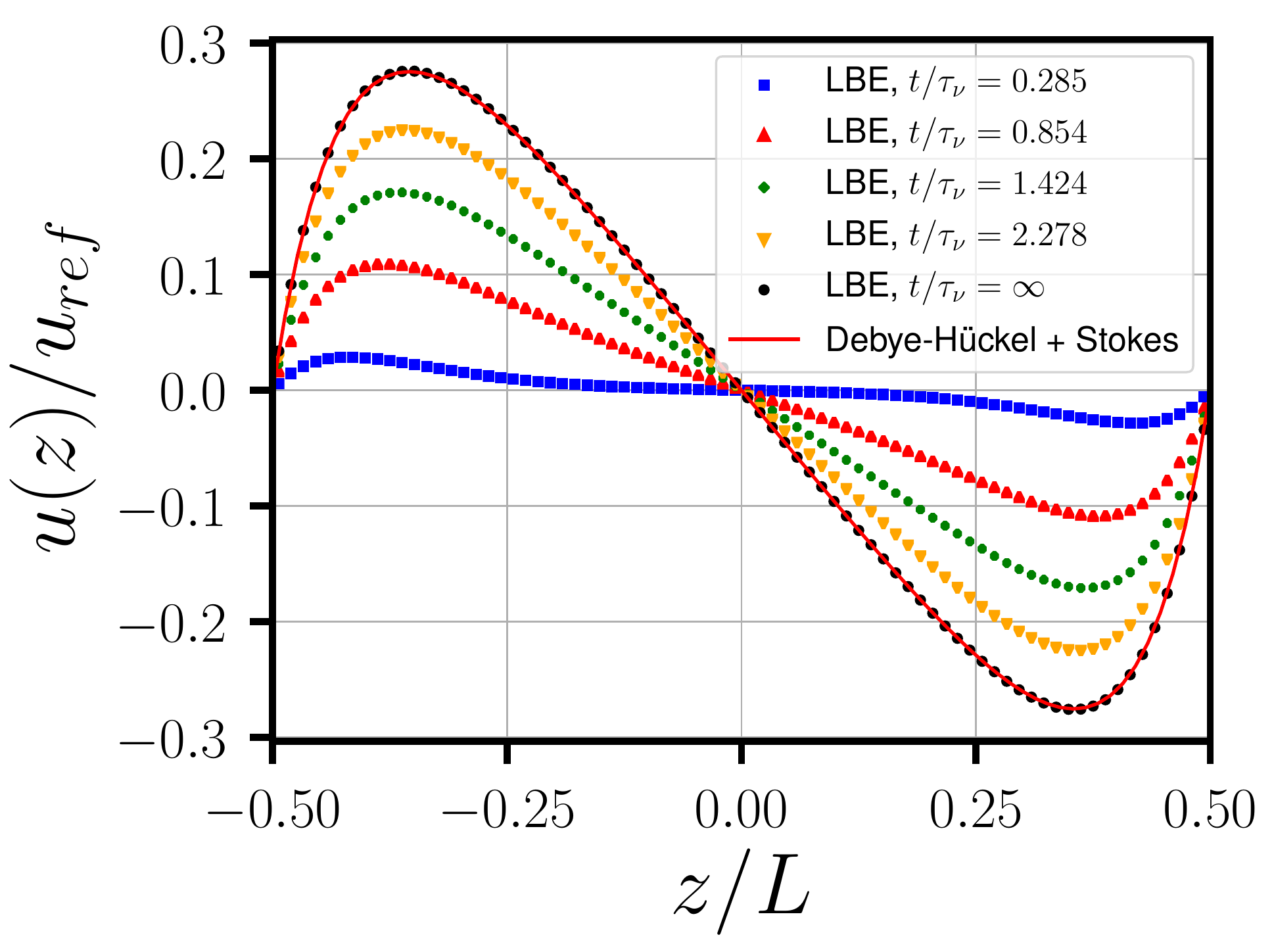}
\caption{
Electro-osmotic flow profile in a charged parallel capacitor,
in the presence of an additional electric field $E_y$ along the electrodes.
The situation at $t=0$ corresponds to the steady state of the charged capacitor.
Lattice-Boltzmann Electrokinetics simulations (LBE) are shown with the symbols.
The steady-state profile is compared to the theoretical result 
Eq.~(\ref{eq:DHslitEOF}) combining 
Debye-H\"uckel theory for the electrostatic potential and the Stokes equation
for the flow (line). Results are scaled with the reference velocity
$u_{ref}=\epsilon_0\epsilon_r E_y (\psi_2 - \psi_1)/\eta$.
Simulations are performed under the same conditions as in
Figure~\ref{fig:slit_phi_rho}, with a reduced applied field $\beta e E
\lbx=0.01$ parallel to the electrodes.
The LBE simulations provide the time-dependence of the electrokinetic response,
which reaches steady-state over a time scale $\tau_\nu=L^2/\pi^2\nu$
with $\nu=\eta/\rho$ the kinematic viscosity of the fluid, as expected
from momentum diffusion in the direction perpendicular to the flow.
}
\label{fig:slit_eof}
\end{figure}

Figure~\ref{fig:slit_eof} reports the simulation results 
corresponding to the system already shown in Figure~\ref{fig:slit_phi_rho}
with an applied electric field in the $y$ direction 
of magnitude $\beta e E_y \lbx=0.01$.
It perfectly reproduces the analytical result expected to be valid
for the considered range of physical parameters, which confirms
the validity of the LBE scheme.
We note that the resulting flow profile corresponds to 
shearing the fluid by applying opposite forces in the two double layers
(since they are oppositely charged). This differs from the common 
situation of shear induced by moving walls in opposite directions, since
the electrodes are not mobile in the present case.

This figure also shows electro-osmotic flow profiles in the transient regime.
The flow builds up in the electric double layers near the electrodes
and develops by momentum diffusion in the direction perpendicular to the
electrodes, over a characteristic time scale $\tau_\nu=L^2/\pi^2\nu$
with $\nu=\eta/\rho$ the kinematic viscosity of the fluid.

As a final remark on the parallel plate capacitor, 
we emphasize again that the comparison is made here only
in the linear regime where DH theory applies
for validation purposes, but that the LBE simulations would 
provide the numerical solution of the non-linear PNP
and Navier-Stokes outside of this regime.

%%%%%%%%%%%
\subsection{Cylindrical (coaxial) capacitor}

\begin{figure}[!htb]
\center
\includegraphics[width=0.45\textwidth]{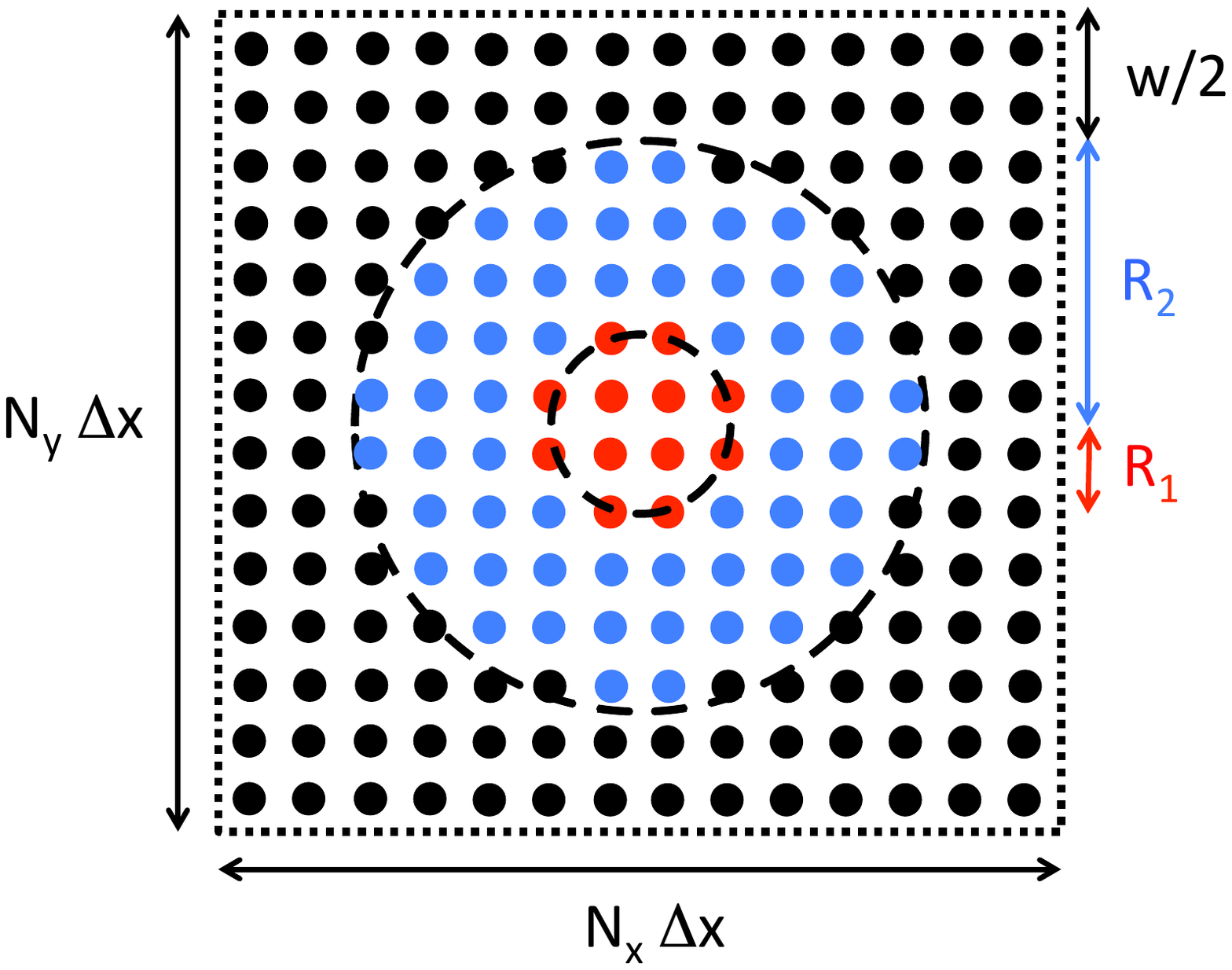}
\caption{
Simulation setup for the coaxial capacitor.
The lattice consists of $N_x\times N_y\times N_z$ nodes
with periodic boundary conditions in all directions
(here a cut in the $xy$ plane is shown) at the boundaries
of the box shown in dotted lines.
The nodes corresponding to the fluid region, illustrated
in blue, are located between two cylinders of radii 
$R_1$ (inner electrode, in red) and $R_2$ (outer electrode, in black)
with potentials $\psi_1$ and $\psi_2$, respectively.
All the region beyond the outer cylinder is maintained at the
same potential (this defines the width $w$ of the electrode region
as shown in the figure).
}
\label{fig:setup_coaxial}
\end{figure}

The setup to simulate cylindrical capacitors is illustrated
in Figure~\ref{fig:setup_coaxial}.
As for the parallel plate geometry, periodic boundary conditions
along $z$ allow in principle to use a single lattice
node in this direction to simulate an infinite system.

%%%%%
\subsubsection{Potential profile}
\label{sec:coaxial_phi}

As for the parallel plate capacitor, we first examine the initial
and steady-state potential profiles within the electrolyte.
LBE simulations were performed in the setup illustrated
in Figure~\ref{fig:setup_coaxial}, with a grid of
$N_x\times N_y\times N_z = 74\times74\times3$ nodes,
a lattice spacing $\lbx=l_B/1.2$,
inner and outer cylinder radii of $R_1=2\lbx\approx1.2$~nm and
$R_2=35\lbx\approx20.4$~nm, 
and a salt concentration ($\approx0.0034$~mol.L$^{-1}$)
corresponding to a screening length
$\lambda_D=9\lbx=7.5l_B\approx5.25$~nm.
With this choice of box size and outer radii, the
width of the outer electrode region is $w=4\lbx$,
see Figure~\ref{fig:setup_coaxial}.

\begin{figure}[!htb]
\center
\includegraphics[width=0.45\textwidth]{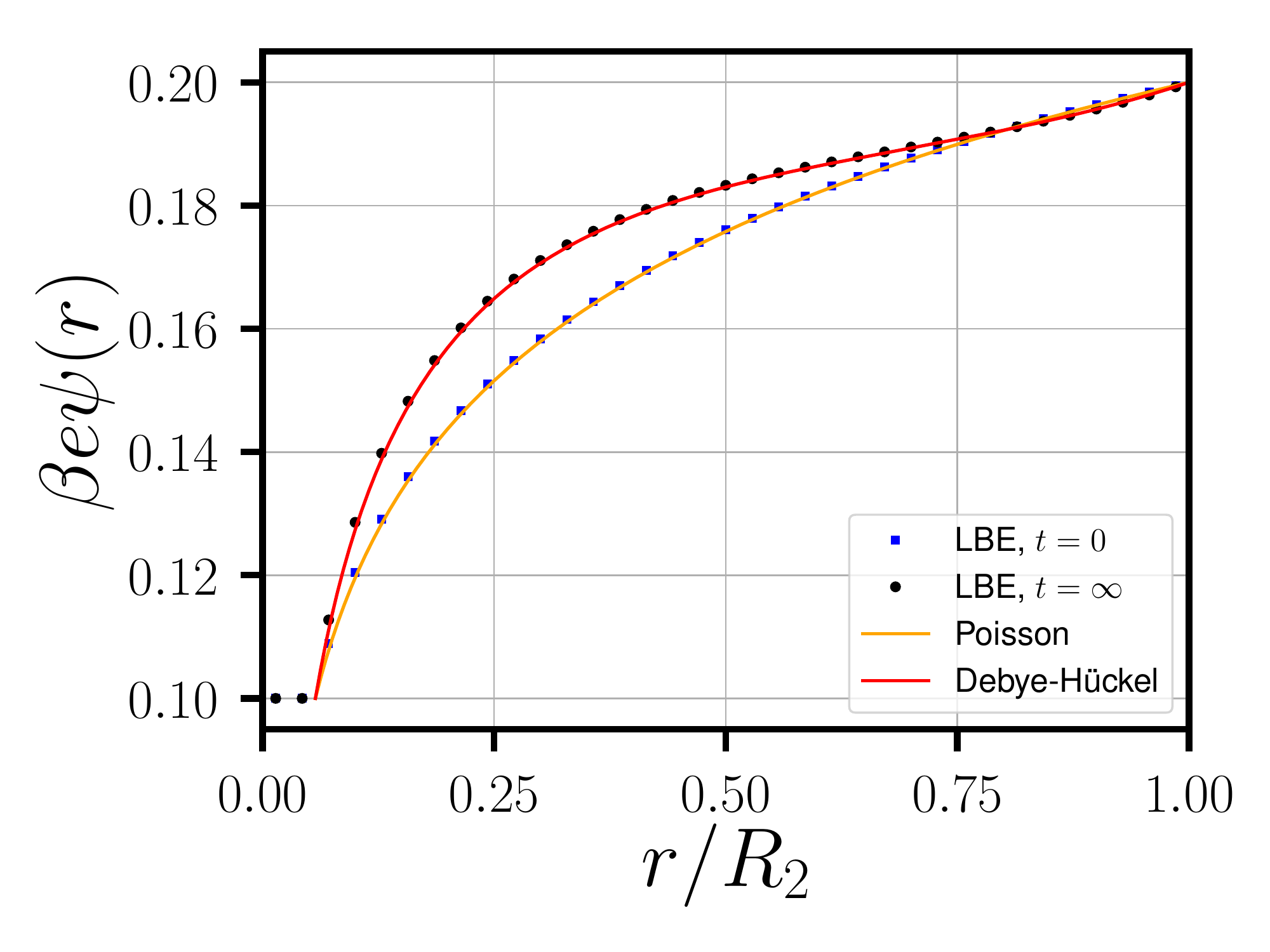}
\caption{
Electrostatic potential profile in a coaxial cylindrical channel,
obtained from Lattice Boltzmann Electrokinetics simulations (LBE, symbols)
and Debye-H\"uckel theory (line).
We also indicate the initial potential profile:
Before the ions start to move, the fluid is neutral and the solution
of the Poisson equation in this geometry is the same
as the one for a simple dielectric coaxial capacitor (see Eq.(\ref{eq:PSIcoxialNosalt}).
Simulations are performed for an inner radius $R_1=2\lbx$ 
and an outer radius $R_2=35\lbx$, with a lattice spacing
$\lbx=l_B/1.2$, with $l_B$ the Bjerrum length, a salt concentration
corresponding to a Debye length $\lambda_D=9\lbx$,
and a reduced voltage $\beta e\Delta\psi=0.1$
between the inner and outer electrodes.
}
\label{fig:coaxial_phi}
\end{figure}

The potential
satisfies the Poisson equation~\ref{eq:Poisson},
with boundary conditions $\psi(R_1)=\psi_1$ and $\psi(R_2)=\psi_2$ 
as well as the constraint of opposite surface 
charge of the two cylinders leading to 
$R_1\psi'(R_1)=R_2\psi'(R_2)$. Before the ions start moving ($t=0$), 
the solution reads:
\begin{align}
\label{eq:PSIcoxialNosalt}
\psi_0^{cyl}(r) &= \psi_1 + (\psi_2-\psi_1)\frac{\ln(r/R_1)}{\ln(R_2/R_1)} 
\end{align}
with $r$ the radial distance from the axis of both cylindrical electrodes.
Figure~\ref{fig:coaxial_phi} shows that the initial potential profile obtained 
numerically with the SOR algorithm is in excellent agreement with this
analytical solution, even though the inner cylinder is discretized
quite roughly ($R_1=2\lbx$ only). This further demonstrates the accuracy
of our numerical scheme to impose constant-potential boundary conditions
in a more complex geometry than planar electrodes.

Figure~\ref{fig:coaxial_phi} also compares the LBE simulation results
for the steady-state potential profile with the analytical
solution of the DH equation~\ref{eq:DH} given by:
\begin{align}
\label{eq:PSIcoxialDH}
\psi_{DH}^{cyl}(r) &= \psi_1 + (\psi_2-\psi_1)f(r,R_1,R_2,\kappa) 
\end{align}
with:
\begin{widetext}
\begin{align}
\label{eq:PSIcoxialDH2}
f(r,R_1,R_2,\kappa) &= 
\frac{  
\left[ R_2 \Bessel{K}{1}{R_2} - R_1 \Bessel{K}{1}{R_1} \right]
\left[ \Bessel{I}{0}{r}-\Bessel{I}{0}{R_1} \right]
+ 
\left[ R_2 \Bessel{I}{1}{R_2} - R_1 \Bessel{I}{1}{R_1} \right]
\left[ \Bessel{K}{0}{r}-\Bessel{K}{0}{R_1} \right]
}{  
\left[ R_2 \Bessel{K}{1}{R_2} - R_1 \Bessel{K}{1}{R_1} \right]
\left[ \Bessel{I}{0}{R_2}-\Bessel{I}{0}{R_1} \right]
 + 
\left[ R_2 \Bessel{I}{1}{R_2} - R_1 \Bessel{I}{1}{R_1} \right]
\left[ \Bessel{K}{0}{R_2}-\Bessel{K}{0}{R_1} \right]
}
\end{align}
\end{widetext}
where $I_{\alpha}$ and $K_{\alpha}$ are modified Bessel functions
of the first and second kind.  
The LBE results are again in excellent agreement with the analytical DH
predictions, which are expected to be valid in this low-voltage regime.

%%%%%
\subsubsection{Capacitance}
\label{sec:coaxial_capa}

We now turn again to the charge induced on the electrode
and corresponding capacitance. The electrode charge per unit
length is coisnveniently derived using Gauss theorem from
the electric field at the surface of the electrodes.
Taking derivatives of the potential with respect to
voltage $\psi_2-\psi_1$ and to the radial distance $r$
(evaluated at $r=R_1$),
it follows from Eqs.~\ref{eq:PSIcoxialNosalt}
and~\ref{eq:PSIcoxialDH}-\ref{eq:PSIcoxialDH2}
that the capacitances per unit length are:
$C_0^{cyl}=2\pi\epsilon_0\epsilon_r/\ln(R_2/R_1)$
for a neutral liquid (before the ions start moving)
and:
\begin{align}
\label{eq:CYLcapaDH}
C_{DH}^{cyl} &= 
2\pi\epsilon_0\epsilon_r R_1 f'(R_1,R_1,R_2,\kappa)
\end{align}
at steady-state (within the Debye-H\"uckel limit).

LBE simulations were performed in the setup illustrated
in Figure~\ref{fig:setup_coaxial}, with a grid of
$N_x\times N_y\times N_z = 54\times54\times3$ node,
inner and outer cylinder radii of $R_1=2\lbx$ and
$R_2=25\lbx$, with a lattice spacing $\lbx=l_B/1.2$.
The reduced potential difference is again fixed to 
$\beta e \Delta \psi=0.1$ and the concentration is varied
over a range corresponding to $\lambda_D/\lbx=3$, 6, 9 and 12.

\begin{table}[!htb]
\center
\begin{tabular}{|*{5}{c}|}
\hline
 $\lambda_D/\Delta x$ & $ 3$ & $ 6$ & $ 9$ & $ 12$   \\
 \hline
$|C_{LBE}-C_{DH}^{cyl}|/C_{DH}^{cyl}$ & 2.3\% &1.2\% & 1.0\% & 0.94\%  \\
\hline
\end{tabular}
\caption{
Relative error on the capacitance, computed at steady-state,
with respect to the theoretical result Eq.~\ref{eq:CYLcapaDH}
in the Debye-H\"uckel limit, for a coaxial capacitor
(see text for simulation details).
}
\label{tab:CapaErrorCoaxial}
\end{table} 

Table~\ref{tab:CapaErrorCoaxial} reports the relative errors for the
capacitance computed at steady-state in the LBE simulations with 
respect to the Debye-H\"uckel analytical result~\ref{eq:CYLcapaDH}
which is expected to be valid in this low-voltage regime.
The errors are very small for the chosen range of simulation parameters.
Similarly to the slit case, the error decreases as $(\lbx/\lambda_D)^2$
when the resolution of the double layer increases.
However, the extrapolated value for $\lbx/\lambda_D\to0$ does not
vanish in that case: This residual value ($\sim0.8\%$) reflects other sources
of errors, in particular due to the coarse discretization of the inner  
cylinder with a radius of only $R_1=2\lbx$.

%%%%%%%%%%%
\subsubsection{Electrokinetic effects}
\label{sec:coaxial_eof}

We finally examine the electrokinetic response of the charged
coaxial capacitor to an additional electric field in the 
axial $z$ direction. The steady-state electro-osmotic flow profile
can be derived from the Stokes equation using the steady-state
potential profile, in the Debye-H\"uckel limit.
The result for no-slip boundary conditions at the surface of the
electrodes reads: 
\begin{align}
\label{eq:DHcoaxialEOF}
u_z(r)&= \frac{\epsilon_0\epsilon_r E_z
(\psi_2 - \psi_1)}{\eta}
\left[
f(r,R_1,R_2,\kappa) - \frac{\ln(r/R_1)}{\ln(R_2/R_1)}
\right]
\end{align}
with $f$ given by Eq.~\ref{eq:PSIcoxialDH2}.

\begin{figure}[!htb]
\center
\includegraphics[width=0.45\textwidth]{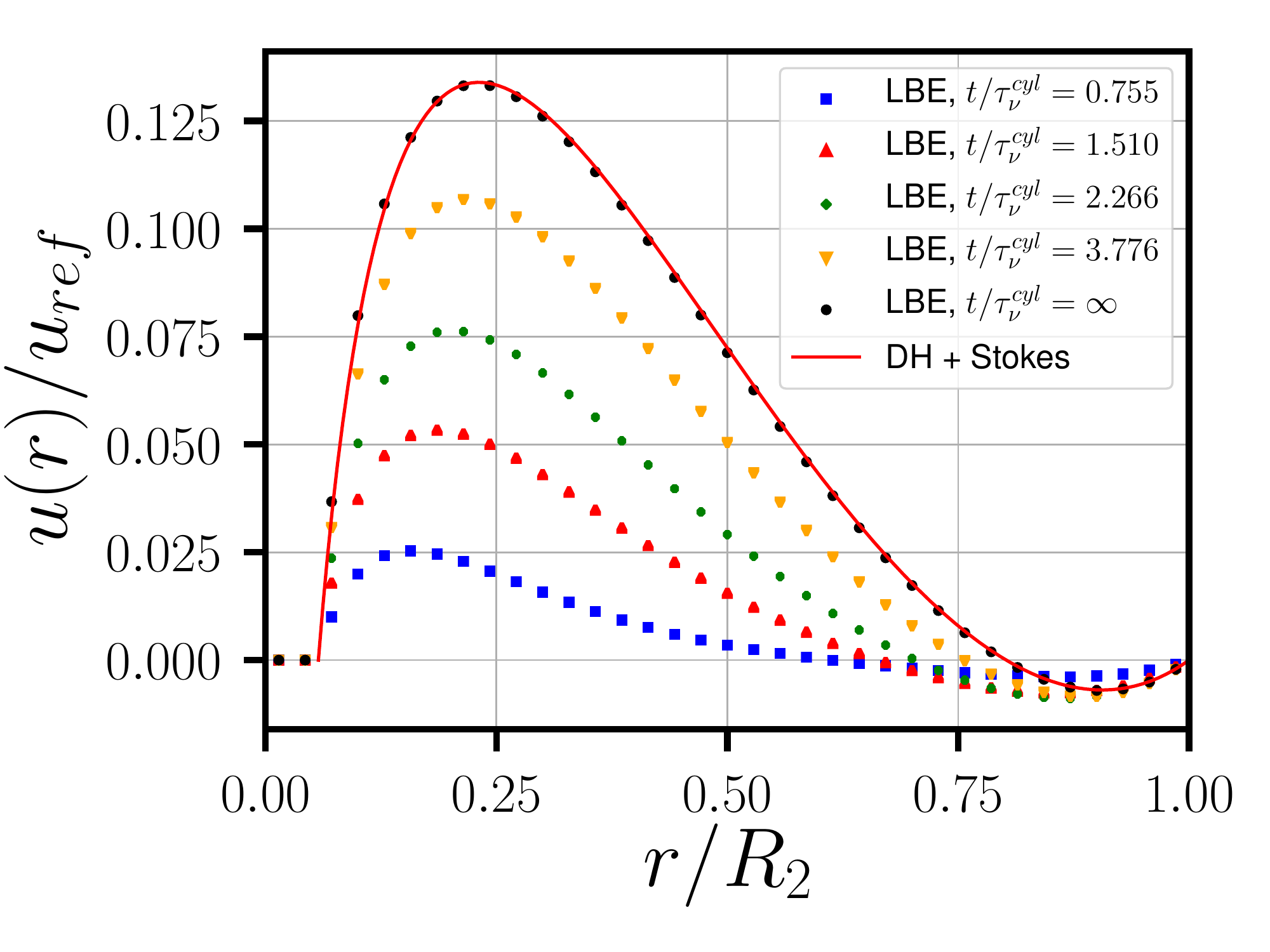}
\caption{
Electro-osmotic flow profile in a coaxial capacitor,
in the presence of an additional electric field $E_z$ along the electrodes.
The situation at $t=0$ corresponds to the steady state of the charged capacitor.
Lattice-Boltzmann Electrokinetics simulations (LBE, symbols)
for the steady-state are compared to the theoretical result Eq.~(\ref{eq:DHcoaxialEOF}) combining 
Debye-H\"uckel theory for the electrostatic potential and the Stokes equation
for the flow (line). Results are scaled with the reference velocity
$u_{ref}=\epsilon_0\epsilon_r E_z (\psi_2 - \psi_1)/\eta$
Simulations are performed under the same conditions as in
Figure~\ref{fig:coaxial_phi}, with a reduced applied field 
$\beta e E_z \lbx=0.0001$ parallel to the electrodes.
The LBE simulations further provide the time-dependence of the electrokinetic 
response, which reaches steady-state over a time scale 
$\tau_\nu^{cyl}=(R_2-R_1)^2/\pi^2\nu$. \cite{comment_tau_cyl}
}
\label{fig:coaxial_eof}
\end{figure}

We performed LBE simulations with the same
parameters as described in section~\ref{sec:coaxial_phi}
for the potential profile. Starting
from the charged capacitor, we apply a reduced electric field
$\beta e E_z \lbx=0.0001$ parallel to the electrodes
(axial direction $z$) and monitor the velocity of the fluid 
in this direction, as a function of radial position $r$
and time $t$. The results shown in Figure~\ref{fig:coaxial_eof}
demonstrate that the steady-state velocity profile is
in excellent agreement with the analytical result 
Eq.~(\ref{eq:DHcoaxialEOF}), as a last illustration
of the validity of the proposed method to impose constant-potential
boundary conditions.
The transient regime (for which no analytical result is available) 
is consistent with the expected acceleration near the electrode
surfaces, where the fluid is not neutral, followed by viscous
momentum diffusion away from these regions to the whole fluid
with a characteristic time $\propto(R_2-R_1)^2/\nu$. \cite{comment_tau_cyl}

As for the parallel plate capacitor, we note that the steady state corresponds 
to shearing the fluid via opposite forces within the two double layers.
This results in particular in flows in opposite directions near the two
electrodes, but with very different magnitudes in that case (larger velocity
near the inner electrode) since the total fluid flux vanishes (there is no
net force on the fluid which is overall neutral).
Such an original setup may find applications to separate species
in a mixture of ions.

%%%%%%%%%%%%%%%%%%%%%%%%%%%%%%%%%%%%%%%%%%%%%%%%%%%%%%%%%%%%%%%%%%%%%%%%%%
%%%%%%%%%%%%%%%%%%%%%%%%%%%%%%%%%%%%%%%%%%%%%%%%%%%%%%%%%%%%%%%%%%%%%%%%%%
\section{Conclusion}

We have introduced a simple rule to impose Dirichlet electrostatic boundary 
conditions in LBE simulations, in a consistent way with the location of
the hydrodynamic interface (for stick boundary conditions), \textit{i.e.}
between the solid and liquid nodes rather than on the solid nodes.
The proposed method also provides the instantaneous local charge 
induced on the electrode by the instantaneous distribution of ions under
voltage. We validated it in the low voltage regime by comparison with
analytical results in two model capacitors (parallel plate and coaxial
electrodes), examining the steady-state ionic concentrations and electric 
potential profiles, the time-dependent response of the charge on the electrodes,
as well as the steady-state electro-osmotic profiles in the presence of an additional, 
tangential electric field. The LBE method naturally provides the time-dependence
of all these quantities -- a possibility that we illustrate on the
electro-osmotic response.
While we do not consider this case in the present work, which focuses
on the validation of the method, the latter readily applies 
to large voltages between the electrodes, as well as to time-dependent
voltages. The only restriction is a sufficiently small lattice spacing, 
with small potential differences (compared to $k_BT/e$) between neighboring nodes.
Besides, we have shown that the method is accurate to second order 
in lattice spacing.

This work opens the way to the LBE simulation of more complex systems
involving electrodes and metallic surfaces, such as the nanofluidic channels 
and nanotubes mentioned in the introduction, or porous electrodes,
since the algorithm can readily be applied to arbitrary geometries. 
It would also be a convenient tool for the simulation of other electrokinetic
phenomena, such as induced-charged electrokinetics\cite{bazant_induced-charge_2010}.
On the methodological side, possible extensions include the
coupling of electrokinetics to adsorption/desorption at the solid-liquid
interface\cite{levesque_accounting_2013,
vanson_unexpected_2015,asta_moment_2018}, which may play a role
in the specific behavior of carbon vs boron nitride
nanotubes\cite{grosjean_chemisorption_2016},
as well as including additional excess terms in the free energy model underlying 
the present work (which only leads to the emergence of the Nernst-Planck dynamics
for the ions). In particular, capturing the effect of ion
correlations\cite{storey_effects_2012} would be necessary to
simulate more concentrated electrolytes as well as multivalent ions.
Finally, it would be useful to obtain analytical results in the non-linear
regime, at least in simple geometries, in order to validate the numerical method
outside of the range considered here. Work in this direction is in progress.

%%%%%%%%%%%%%%%%%%%%%%%%%%%%%%%%%%%%%%%%%%%%%%%%%%%%%%%%%%%%%%%%%%%%%%%%%%
%%%%%%%%%%%%%%%%%%%%%%%%%%%%%%%%%%%%%%%%%%%%%%%%%%%%%%%%%%%%%%%%%%%%%%%%%%
%%%%%%%%%%%%%%%%%%%%%%%%%%%%%%%%%%%%%%%%%%%%%%%%%%%%%%%%%%%%%%%%%%%%%%%%%%
%\bibliography{BibliographyThesis}

\begin{thebibliography}{63}
\bibitem [{com()}]{comment_tau_cyl}%
  \BibitemOpen
  \href@noop {} {}\bibinfo {note} {The exact expression for the largest
  momentum relaxation time is
  $\tau_\nu^{cyl}/\alpha^2=(R_2-R_1)^2/\alpha^2\pi^2\nu$. The value of $\alpha$
  is determined by the following equation: $ J_0(\frac{\pi \alpha
  R_1}{R_2-R_1}) Y_0(\frac{\pi \alpha R_2}{R_2-R_1}) = J_0(\frac{\pi \alpha
  R_2}{R_2-R_1}) Y_0(\frac{\pi \alpha R_1}{R_2-R_1}) $, where $J_0$ and $Y_0$
  are Bessel functions of the first and second kind. Here, $\alpha^2$ is 0.867
  and, in general, a numerical solution of the equation shows that
  $0.7<\alpha^2<1$}\BibitemShut {NoStop}%
\end{thebibliography}


%%%%%%%%%%%%%%%%%%%%%%%%%%%%%%%%%%%%%%%%%%%%%%%%%%%%%%%%%%%%%%%%%%%%%%%%%%
\begin{acknowledgments}
The authors are grateful to Lyd\'eric Bocquet and Ignacio Pagonabarraga
for useful discussions.
AJA and BR acknowledge financial support from the French Agence Nationale de
la Recherche (ANR) under grant ANR-15-CE09-0013-01.
The work was funded by the European Union's Horizon 2020 research and innovation
programme under ETN grant 674979-NANOTRANS. 
\end{acknowledgments}

\end{document}